\documentclass[referee,longauth]{aa} 
\usepackage[varg]{txfonts}
\usepackage{graphicx}           
\usepackage{amssymb}            
\usepackage{natbib}             
\bibpunct{(}{)}{;}{a}{}{,}      
\usepackage{float}             

\newcommand{\eq}[1]{Eq.~(\ref{eq:#1})}

\newcommand{\eqs}[2]{Eqs.~(\ref{eq:#1},\ \ref{eq:#2})}
\newcommand{\Eqs}[2]{Equations~(\ref{eq:#1},\ \ref{eq:#2})}

\newcommand{\sect}[1]{Sect.~\ref{s:#1}}
\newcommand{\sects}[2]{Sects.~\ref{s:#1} and \ref{s:#2}}

\newcommand{\fig}[1]{Fig.~\ref{fig:#1}}
\newcommand{\Fig}[1]{Figure~\ref{fig:#1}}
\newcommand{\Figs}[1]{Figures~\ref{fig:#1}}

\newcommand{\BE}{\begin{equation}}
\newcommand{\EE}{\end{equation}}
\newcommand{\BA}{\arraycolsep=1.0pt\begin{eqnarray}}
\newcommand{\EA}{\end{eqnarray}}
\newcommand{\BI}{\begin{itemize}}
\newcommand{\EI}{\end{itemize}}

\newcommand{\eg}{\textit{e.g.},}
\newcommand{\ie}{\textit{i.e.},}

\newcommand{\SDO}{SDO}
\newcommand{\HMI}{\SDO{}/HMI}
\newcommand{\hmiBlos}{\texttt{hmi.m\_45s}} 
\newcommand{\hmiME}{\texttt{hmi.ME\_720s\_fd10}}
\newcommand{\SolO}{Solar Orbiter}
\newcommand{\SO}{SO}
\newcommand{\PHI}{{SO/PHI}}
\newcommand{\HRT}{\PHI{}-HRT}

\newcommand{\PaperI}{Paper~I}
\newcommand{\direct}{``direct''}   
\newcommand{\reverse}{``reverse''}  
\newcommand{\SDM}{SDM}
\newcommand{\ME}{ME0}
\newcommand{\vn}{\vec{n}}
\newcommand{\vv}{\vec{v}}
\newcommand{\vl}{\vec{l}}

\newcommand{\vw}{\vec{w}}

\newcommand{\hatn}{\hat{\vv}}
\newcommand{\hatl}{\hat{\vl}}

\newcommand{\hatw}{\hat{\vw}}

\newcommand{\lA}{\hatl_{\rm A}}
\newcommand{\wA}{\hatw_{\rm A}}

\newcommand{\lB}{\hatl_{\rm B}}
\newcommand{\wB}{\hatw_{\rm B}}
\newcommand{\SA}{S_{\rm A}}
\newcommand{\SB}{S_{\rm B}}
\newcommand{\sgn}{\zeta}

\newcommand{\vB}{\vec{B}}

\newcommand{\Btr}{B_{\rm tr}}
\newcommand{\Blos}{B_{\rm los}}
\newcommand{\BlA} {\Blos^{\rm A}}

\newcommand{\BtrA}{\Btr^{\rm A}}
\newcommand{\BwA} {B_{\rm w}^{\rm A}}
\newcommand{\BnA} {B_{\rm v}^{\rm A}} 
\newcommand{\alA} {\alpha^{\rm A}}
\newcommand{\thA} {\theta^{\rm A}}

\newcommand{\BlB} {B_{\rm los}^{\rm B}}
\newcommand{\BtrB}{\Btr^{\rm B}}
\newcommand{\BwB} {B_{\rm w}^{\rm B}}
\newcommand{\BnB} {B_{\rm v}^{\rm B}}
\newcommand{\alB} {\alpha^{\rm B}}
\newcommand{\thB} {\theta^{\rm B}}

\newcommand{\Bw} {B_{\rm w}}
\newcommand{\Bn} {B_{\rm v}}
\newcommand{\dBn} {\delta\Bn}
\newcommand{\PIL} {PIL$_{\rm{LoS}}$}

\newlength{\imsize}
\newlength{\labshift}

\newcommand{\labfigure}[5]{              
  \setlength{\labshift}{#3}
  \begin{minipage}[t]{#1}
    \includegraphics[width= #1 ]{#2}
    \hspace{- #1 } 
    \hspace{\labshift} 
    \raisebox{#4}{\textbf{#5}}
  \end{minipage}
 }
\newcommand{\labfigurecr}[9]{
  \setlength{\labshift}{#3}
  \begin{minipage}[t]{#1}
    \includegraphics[width= #1 , clip= , trim= #6 #7 #8 #9]{#2}
    \hspace{- #1 }
    \hspace{\labshift}
    \raisebox{#4}{\textbf{#5}}
  \end{minipage}
 }

\newcommand{\locpath}{./}
\newcommand{\figpath}{\locpath}           
\newcommand{\figME}{\figpath}
\newcommand{\figBlos}{\figpath}  

\usepackage[normalem]{ulem}

\newcommand{\kd}[1]{{#1}}
\newcommand{\kdtwo}[1]{{#1}}         

\begin{document}
\title{Stereoscopic disambiguation of vector magnetograms: \\ first applications to \HRT{} data}
\titlerunning{Stereoscopic disambiguation}
\authorrunning{G.~Valori et al.}
\author{
   G.~Valori \inst{1} \orcid{0000-0001-7809-0067}
   \thanks{Corresponding author: G. Valori \email{valori@mps.mpg.de}} \and
   D.~Calchetti\inst{1}\orcid{0000-0003-2755-5295} \and         
   A.~Moreno Vacas\inst{2}\orcid{0000-0002-7336-0926} \and      
   \'E.~Pariat\inst{3}\orcid{0000-0002-2900-0608} \and          
   S.K.~Solanki\inst{1}\orcid{0000-0002-3418-8449} \and         
   P.~L\"oschl\inst{1}\orcid{0000-0002-0038-7968} \and          
   J.~Hirzberger\inst{1} \and                                   
   S.~Parenti\inst{4}\orcid{0000-0003-1438-1310} \and
   K.~Albert\inst{1}\orcid{0000-0002-3776-9548} \and            
   N. Albelo~Jorge\inst{1} \and                                 
   A.~\'Alvarez-Herrero\inst{5}\orcid{0000-0001-9228-3412} \and 
   T.~Appourchaux\inst{4} \and                                  
   L.R.~Bellot~Rubio\inst{2}\orcid{0000-0001-8669-8857} \and    
   J.~Blanco~Rodr\'\i guez\inst{6}\orcid{0000-0002-2055-441X}\and
   A.~Campos-Jara\inst{5}\orcid{0000-0003-0084-4812}            
   A.~Feller\inst{1} \and                                       
   A.~Gandorfer\inst{1}\orcid{0000-0002-9972-9840} \and         
   P.~Garc\'\i a~Parejo\inst{5}\orcid{0000-0003-1556-9411} \and 
   D.~Germerott\inst{1} \and                                    
   L.~Gizon\inst{1,11}\orcid{0000-0001-7696-8665} \and          
   J.M.~Gómez~Cama\inst{8}\orcid{0000-0003-0173-5888} \and      
   L.~Guerrero\inst{1} \and                                     
   P.~Gutierrez-Marques\inst{1}\orcid{0000-0003-2797-0392} \and 
   F. Kahil\inst{1}\orcid{0000-0002-4796-9527} \and             
   M.~Kolleck\inst{1} \and                                      
   A.~Korpi-Lagg\inst{1}\orcid{0000-0003-1459-7074} \and        
   D.~Orozco~Su\' arez\inst{2}\orcid{0000-0001-8829-1938} \and  
   I.~P\' erez-Grande\inst{9}\orcid{0000-0002-7145-2835} \and   
   E.~Sanchis Kilders\inst{6}\orcid{0000-0002-4208-3575} \and   
   J.~Schou\inst{1}\orcid{0000-0002-2391-6156} \and             
   U.~Sch\" uhle\inst{1}\orcid{0000-0001-6060-9078} \and       
   J.~Sinjan\inst{1}\orcid{0000-0002-5387-636X} \and            
   J.~Staub\inst{1}\orcid{0000-0001-9358-5834} \and             
   H.~Strecker\inst{2}\orcid{0000-0003-1483-4535} \and          
   J.C.~del~Toro~Iniesta\inst{2}\orcid{0000-0002-3387-026X} \and
   R.~Volkmer\inst{10}\and                                      
   J.~Woch\inst{1}\orcid{0000-0001-5833-3738}                   
}     
     
   \institute{
         Max-Planck-Institut f\"ur Sonnensystemforschung, Justus-von-Liebig-Weg 3,
         37077 G\"ottingen, Germany \\ \email{solanki@mps.mpg.de}
         \and
         Instituto de Astrofísica de Andalucía (IAA-CSIC), Apartado de Correos 3004,
         E-18080 Granada, Spain \\ \email{jti@iaa.es}
         \and
         Sorbonne Universit\'e, \'Ecole polytechnique, Institut Polytechnique de Paris, Universit\'e Paris Saclay, Observatoire de Paris-PSL, CNRS, Laboratoire de Physique des Plasmas (LPP), 75005 Paris, France  
         \and
         Universit\'e Paris-Saclay, CNRS, Institut d’Astrophysique Spatiale, 91405, Orsay, France
         \and
         Instituto Nacional de T\' ecnica Aeroespacial, Carretera de
         Ajalvir, km 4, E-28850 Torrej\' on de Ardoz, Spain
         \and
         Universitat de Val\`encia, Catedr\'atico Jos\'e Beltr\'an 2, E-46980 Paterna-Valencia, Spain
         \and
         Institut f\"ur Datentechnik und Kommunikationsnetze der TU
         Braunschweig, Hans-Sommer-Str. 66, 38106 Braunschweig,
         Germany
         \and
         University of Barcelona, Department of Electronics, Carrer de Mart\'\i\ i Franqu\`es, 1 - 11, 08028 Barcelona, Spain
         \and
         Instituto Universitario "Ignacio da Riva", Universidad Polit\'ecnica de Madrid, IDR/UPM, Plaza Cardenal Cisneros 3, E-28040 Madrid, Spain
         \and
         Leibniz-Institut für Sonnenphysik, Sch\" oneckstr. 6, D-79104 Freiburg, Germany
         \and
         Institut f\"ur Astrophysik, Georg-August-Universit\"at G\"ottingen, Friedrich-Hund-Platz 1, 37077 G\"ottingen, Germany
         }

\date{Received ***; accepted ***}

   \abstract
{
Spectropolarimetric reconstructions of the photospheric vector magnetic field are intrinsically limited by the 180$^\circ$-ambiguity in the orientation of the transverse component. 
So far, the removal of such an ambiguity has required assumptions about the properties of the photospheric field, which makes disambiguation methods model-dependent.   
}
{
The successful launch and operation of Solar Orbiter have made the removal of the 180$^\circ$-ambiguity possible solely using observations of the same location on the Sun obtained from two different vantage points. 
}
{
The basic idea is that the unambiguous line-of-sight component of the field measured from one vantage point will generally have a non-zero projection on the ambiguous transverse component measured by the second telescope, thereby determining the ``true'' orientation of the transverse field.
Such an idea was developed and implemented in the Stereoscopic Disambiguation Method (SDM), which was recently tested using numerical simulations.
}
{
In this work we present a first application of the SDM to  data obtained by the High Resolution Telescope (HRT) onboard Solar Orbiter during the March 2022 campaign, when the angle with Earth  was 27 degrees. 
The method is successfully applied to remove the ambiguity in the transverse component of the vector magnetogram solely using observations (from HRT and from the Helioseismic and Magnetic Imager), for the first time. 
}
{
The SDM is proven to provide observation-only disambiguated vector magnetograms that are spatially homogeneous and consistent.  
A discussion about the sources of error that may limit the accuracy of the method, and of the strategies to remove them in future applications, is also presented.
}

   \maketitle

\section{Introduction}\label{s:intro}
The \kd{solar} photospheric magnetic field can be inferred from spectropolarimetric observations by parametrically matching the measured Stokes spectra with synthetic profiles based on radiative transfer in atmospheric models \citep[see, \eg{}][]{Lites2000}. 
Such a technique (called inversion) can only provide a partial knowledge of the vector field, namely the amplitude and orientation of the field component along the line of sight (LoS hereafter) of the observer, and the amplitude and direction of the field component perpendicular to the LoS.
However, no information is provided about the orientation along the transverse direction of the field, resulting in an ambiguity of the transverse component: two orientations of the transverse component that differ by 180$^\circ$ are indistinguishable from each other.  
Such an ambiguity is due to the invariance of the Stokes vector to a 180$^\circ$ rotation of the reference system about the LoS-axis. 
Therefore, the 180$^\circ$ ambiguity in the orientation of the transverse field component is an intrinsic limitation of remote sensing that cannot be eliminated by improving spectropolarimetric measurements.
On the other hand, solving the 180$^\circ$ ambiguity (disambiguation) corresponds to the determination of the sign of the transverse component in each pixel of the image plane, and it is therefore a parity problem \citep{Semel1998}.

The importance of a correct disambiguation of vector magnetograms can hardly be overestimated.
Besides its implication for a proper description of the evolution of the photospheric magnetic field, the orientation of the transverse component \kd{enters the computation of} the electric currents that are injected into the upper layers of the solar atmosphere.
\kd{It is only after disambiguation that the observed magnetic field components can be reprojected in the physical radial, toroidal, and poloidal components \citep{Gary1990,Sun2013} that are used to compute physically-relevant quantities such as radial currents.} 
In turn, such currents are the origin of the free magnetic energy that powers coronal activity \citep[see, \eg][]{Forbes2006,Aulanier2013}, from flares to coronal mass ejections, contributing to coronal heating as well as to the variability of the heliospheric environment.

Several empirical methods are available that propose solutions for removing the 180$^\circ$ ambiguity, see for a review \cite{Metcalf2006}, and \cite{Leka2009} for  additional comparisons.
A common limitation of all traditional disambiguation methods is that, in order to compensate for the incomplete information about the transverse magnetic field, they must necessarily rely on assumptions in order to constrain its orientation.
Such assumptions may vary from simply choosing the orientation of the transverse field that is closer to the corresponding potential field, to more complex criteria that involve minimizing a weighted combination of vertical electric currents and divergence of the magnetic field. 
Typically, such criteria are formulated as a minimization problem, and Table~1 in \cite{Leka2009} lists the quantity to be minimized for several methods. 
However, as reasonable as they can be, such simplifying assumptions might not be always fulfilled, especially in highly complex magnetic field regions that are often the source of space-weather relevant flare events. 

A new possibility for the solution of the 180$^\circ$ ambiguity is offered by the successful launch and operation of \SolO{} \citep[\SO{},][]{Mueller2020,Zouganelis2020} and its onboard magnetograph, the Polarimetric and Helioseismic Imager \citep[\PHI{},][]{Solanki2020}.
The orbit of \SO{} allows for remote-sensing observations from different vantage points away from the Sun-Earth line.
The combination of information on the magnetic field orientation from different viewpoints of the same area on the Sun can now be used to remove the ambiguity observationally \citep{Solanki2015,Rouillard2020}. 

\kd
{Regardless of the employed method, either single-view or stereoscopic, disambiguation is a step that is affected by the accuracy of the input information.  
The optical quality of the observations, the limitations and accuracy of the employed inversion technique  deducing the magnetic vector from spectropolarimetric data, and the accuracy of any geometrical transformation that may be needed for the application of the chosen disambiguation method, all such factors may influence the success of the disambiguation. 
Such factors should be regarded as influencing, but not being intrinsically part of, the disambiguation method. 
As mentioned above, single-view methods do require additional hypothesis in order to remove the ambiguity, but have the advantage of dealing with such problems for one instrument only. 
On the other hand, any stereoscopic method is confronted with the additional complication of bringing together information from two different instruments, which prominently entails differences in calibrations, resolution, and generally inversion techniques, in addition to requiring a sensitive geometrical procedure for combining views from different vantage points.}

\cite{Valori2022}, hereafter \PaperI{}, introduced the Stereoscopic Disambiguation Method (\SDM{}) that solves the 180$^\circ$ ambiguity by combining information from two vantage points.
Differently from traditional, single-viewpoint methods, the \SDM{}  solves the ambiguity based on observations only, without any assumption about the magnetic field.
\PaperI{} also used numerically-simulated vector magnetograms to show that the \SDM{} can remove the ambiguity with great accuracy in a large range of stereoscopic angles. 

In this article we present the very first application of the \SDM{} to real observations.
We use co-temporal observations from the High Resolution Telescope \citep[\HRT{},][]{Gandorfer2018} of \PHI{} onboard \SO{}, and from the Helioseismic and Magnetic Imager \citep[\HMI{},][]{Scherrer2012a,Schou2012} onboard the 
Solar Dynamics Observatory
\citep[\SDO{},][]{Pesnell2012}.

%
\kdtwo
{
The \SDM{} and its application are described in \sect{method}. 
In \sect{data} we provide details of the datasets used in the tests presented, both from \HRT{} and from two distinct \HMI{}-data product series. 
The results of two \SDM{} disambiguations of the same \HRT{} vector magnetogram but using the two \HMI{}  data series are presented in \sect{sdm_720_reverse} and \ref{s:sdm_45_reverse_los}. 
In \sect{discussion} we discuss possible sources of error and future improvements of the \SDM{}, and \sect{conclusions} summarizes our conclusions.
}

\section{The stereoscopic disambiguation method}\label{s:method}
The stereoscopic disambiguation of vector magnetograms can be split into two independent steps: first, the solution of the geometrical problem of relating the components of a same vector magnetic field as seen from two different vantage points (\sect{geometry}). 
Second, the practical application of the method to real observations (\sect{application}) and its numerical implementation (\sect{implementation}).  
\subsection{\SDM{}: the geometrical problem}\label{s:geometry}
\PaperI{} derives the equations to determine the sign $\sgn=\pm1$ of the transverse component in each pixel of the image planes of two telescopes.
For each telescope, \eg{} for the telescope A,  a \SDM{} reference system $\SA$ is defined by three unit-vectors: the direction of the LoS ($\lA$), the common normal ($\hatn$) to the plane through the telescopes A and B and the center of the Sun, and their normal vector ($\wA=\hatn\times\lA$, see Fig.~1 in \PaperI{}.
\kd{Please, notice the change in notation $\hat{\vn} \rightarrow \hatn$ with respect to \PaperI{}, adopted here to avoid any possible confusion with the component of the field normal to the solar surface}
).    
In $\SA(\lA,\wA,\hatn)$ the magnetic field, $\vB$,  is written as 
\BE
\vB=\BlA\lA+\sgn\left(\BwA\wA+\BnA\hatn\right),  \label{eq:field} \\
\EE
where $\BlA$ is the (signed) LoS-component and 
\BE
\BwA=\BtrA\cos\alA, \qquad \BnA=\BtrA\sin\alA \label{eq:components}
\EE
with the polar angle $\alA$ defined in $[0,\pi]$ and $\BtrA$ is the (positive-defined) amplitude of the transverse component. 
Analogous expressions hold for the reference system $\SB(\lB,\wB,\hatn)$ of the telescope B, namely
\BE
\vB=\BlB\lB+\sgn\left(\BwB\wB+\BnB\hatn\right),  \label{eq:fieldSB} \\
\EE
with
\BE
\BwB=\BtrB\cos\alB, \qquad \BnB=\BtrB\sin\alB \label{eq:componentsSB}
\EE

In practice, $\SA$ (respectively, $\SB$) is a rotation of the detector reference system by an angle $\thA$ (respectively, $\thB$) around the LoS such that the detector y-direction is parallel to $\hatn$.
Since by construction $\hatn$ is the same for $\SA$ and $\SB$, and since both $\alA$ and $\alB$ are restricted between 0 and $\pi$, it follows that the $\hatn$-components on $\SA$ and $\SB$ are identical regardless of the ambiguity, \ie{} $\BnA=\BnB$ (and the sign function $\sgn$ is the same in \eq{field} and \eq{fieldSB}, see in particular Eq.~(8) in \PaperI{} for a proof). 
The property that  $\BnA=\BnB$ is used in \sect{diag_Bn} as a consistency criteria for the application of the \SDM{}.
    
Using the above representation, \PaperI{} shows that the sign of the transverse component, $\sgn$, is given by either of the two geometrically-equivalent formulae 
\BA
\sgn&=&\frac{\BlA\sin\gamma}{\BwA\cos\gamma-\BwB} \label{eq:sdm_tr}, \\
\sgn&=&\frac{\BlB-\BlA\cos\gamma}{\BwA\sin\gamma} \label{eq:sdm_los},
\EA
where $\gamma$ is the separation angle between the two telescopes A and B, defined as counter-clockwise around $\hatn$ from the direction of telescope A.

In \eq{sdm_tr} the $\Bw$-components of both telescopes A and B appear, while in \eq{sdm_los} only $\BwA$ does. 
In other words, \eq{sdm_los} can be applied to disambiguate the transverse component on telescope A even if only the LoS component on telescope B is available (see \sect{sdm_45_reverse_los} for an application of this particular case).

\subsection{\SDM{}: application to observations}\label{s:application}

\Eqs{sdm_tr}{sdm_los} can be applied in several ways to data from \HRT{} and \HMI{}.
First, one can disambiguate A-magnetograms using information from B (the \direct{} case of Sect~3.3.1 in \PaperI{}), or vice versa (the \reverse{} case).
Second, the sign $\sgn$ of the transverse component can be determined by either of \eqs{sdm_tr}{sdm_los}, or by a combination of both (see Sect.~3.3.2 in \PaperI{} in particular).
Finally, different \HMI{} data series can be used as input for the \SDM{}.

In this paper we consider the disambiguation of the \HRT{} dataset using two different \HMI{} series (\sect{data_hmi}). 
In the terminology of  \PaperI{}, this case corresponds to the \reverse{} application of the \SDM{} with the associations A=\HRT{} and B=\HMI{}.
In this first application we do not consider disambiguation of \HMI{} using \HRT{} because of the unfavorable position of the target active region (AR) on the solar disk (see \sect{data}).

\subsection{\SDM{}: numerical implementation}\label{s:implementation}

In order to use \eqs{sdm_tr}{sdm_los} in real applications, the observed vector magnetic field from each telescope must be first transformed from the image plane 
to the corresponding \SDM{} reference system. 
In this section we provide an example of the \SDM{} workflow for the \reverse{} applications presented in the following sections.
The \SDM{} software is developed using the SolarSoft suite of programs \citep{Freeland2012}.

The following steps are performed to apply the \SDM{} to a given \HRT{} dataset:
\begin{enumerate}
\item \HMI{} data selection. 
The \HMI{} magnetogram should be chosen as the one closest in time to the considered \HRT{} observation.
However, due to the generally smaller distance of \SO{} from the Sun, the difference in light travel-time between \SO{} and \SDO{} must be considered.
Hence, the selected \HMI{} dataset is chosen as the closest in time \kd{\citep[as specified by the \texttt{T\_OBS} FITS keyword, see][]{Couvidat2016}} to the time at which the light measured by \HRT{} would have reached Earth.
The latter is readily provided by the FITS keyword \texttt{DATE\_EAR} in the FITS header of the \HRT{} dataset.
No further time adjustments is considered at this stage.
\item Image co-registration.
This step is crucial for the application of the \SDM{}, which intrinsically requires the alignment between images to have subpixel accuracy.
Currently, the World Coordinate System \kd{\citep[WCS,][]{Thompson2006}} keywords in \HRT{} do not match the \HMI{} ones to such a degree of accuracy, therefore a co-registration step is unavoidable.   
The co-registration is performed using the field strength as comparison when possible, which is ideally independent of the viewing angle, unless only $\Blos$ is available (as in \sect{sdm_45_reverse_los}).

The \HMI{} image is first remapped onto the \HRT{} detector frame, then a subdomain of co-registration is chosen.
\kd{The remapping includes the (bi-linear) interpolation of the \HMI{} image onto the \HRT{} uniform grid.} 
The co-registration technique is a simple but fast and accurate Fourier matching technique that provides the co-registration shifts to apply to the \HRT{} reference pixel identified by the WCS keywords \texttt{CRPIX1} and \texttt{CRPIX2}. 
The remapping/co-registration steps are repeated, each time  using the newly updated \HRT{} WCS keywords, until convergence is reached to the desired precision (to better than 1/100$^{th}$ of a pixel, in the application here). 
We verified that, for the \HRT{} dataset considered here, the correction to the WCS keyword \texttt{CROTA}, representing the rotation angle of the detector with respect to solar north, is negligible. 
The co-registration procedure is in principle independent of the subdomain where the \SDM{} is applied, although here we use the same field of view (FoV) for both.
\item Remapping of the \HMI{} Cartesian magnetic field components on the \HRT{} detector frame.
This step uses the co-registered \HRT{} WCS information as updated in step~2.
The remapping entails \kd{a (bi-linear)} interpolation of the \HMI{} magnetogram (upscaling \HMI{} to the \HRT{} resolution in the cases presented here).
\item Computation of the separation angle $\gamma$, and of the \SDM{} rotation angles $\thA$ and $\thB$. 
These angles are directly computed from the observer Carrington coordinates as included in the FITS headers of the considered \HRT{}  and \HMI{} dataset. 
\item Reprojection of the \HRT{}  and \HMI{} vector magnetic fields on the \SDM{} reference systems, $\SA$ and $\SB$ respectively.
This step produces the representation of the vector magnetic field according to \eqs{field}{fieldSB}, with the azimuth of both fields re-normalized to be within $[0,\pi]$.
\item Application of the relevant \SDM{} equation (either \eq{sdm_tr} or \eq{sdm_los}) to compute the sign function, $\sgn$.
In order to constrain $\sgn$ to the nominal $\pm1$ values, we then build a parity map by taking the sign of $\sgn$ (see also \sect{diag_sgn}). 
\item The parity map is then applied to the \HRT{} transverse component to remove the ambiguity.
Once disambiguated, the \HRT{} field is reprojected back to the detector reference system for ease of comparison.

\end{enumerate}
\section{Observations and data preparation}\label{s:data}
\begin{figure}
 \setlength{\imsize}{\columnwidth}
 \centering
 \includegraphics[width=\imsize,clip=,trim=50 50 20 40]{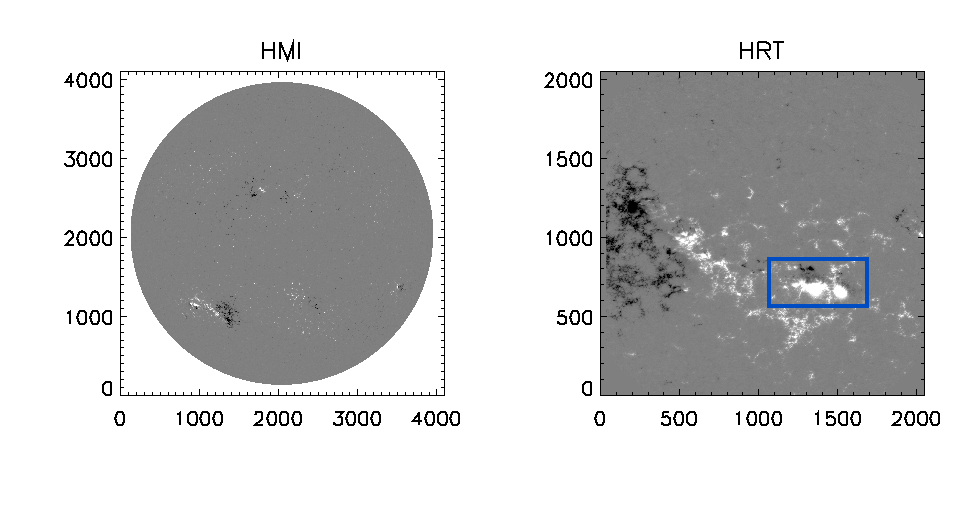}
 \caption{$\Blos$ on 17-Mar-2022 on the image planes of \HMI{} (left, at 03:46:35.5\,UT) and \HRT{} (right, at 03:44:09\,UT). 
          The blue rectangle on the \HRT{} image shows the subdomain that is used for  co-registration (see step~2 in \sect{implementation}) and \SDM{} application.
 The images are not rotated, meaning that solar north is approximately down in \HMI{} (left) and up in \HRT{} (right).
 In both panels, axes are in pixels. 
 }
 \label{fig:fov}
 \end{figure}
We considered the observation of NOAA AR12965 taken by \HRT{} and \HMI{} on March 17th, 2022, when 
\SO{} was at 0.38\,AU from the Sun and the separation angle  with 
\SDO{} was $\gamma=26.53^{\circ}$.
\kdtwo{The Carrington longitude and latitude of the target AR were $(271.5^\circ, 22.5^\circ)$, respectively. }
\Fig{fov} shows the LoS magnetic field ($\Blos$) of the chosen datasets on the full image planes of the two telescopes.
The rationale behind this choice amongst those available in the first months of the science mission-phase of \SO{} is that it contains a well-formed AR within the FoV, and that observations are taken when the two telescopes are well within the range of stereoscopic angles, \ie{} far enough from both quadrature and inferior conjunction (see also Sect.~4.4 in \PaperI{}).
The AR contains two compact sunspots of positive polarities (see also \fig{cnt_720_vect}) close to each other, and a more dispersed, following negative-polarity region. 
The two positive sunspots in the AR are of particular interest for this first application of the \SDM{} because they allow for some qualitative considerations about the expected orientation of the transverse component.
The same AR was also studied in \cite{Li2023}, where observations from the Extreme Ultraviolet Imager \citep{Rochus2020} onboard \SO{} were exploited to study oscillations in coronal loops. 

\kdtwo{
While the above criteria identified the chosen dataset as the only suitable one available at the time of writing, it is still not ideal. 
First, because the AR was relatively close to the limb of \HMI{} (see the left panel in \fig{fov}), it is therefore affected by strong foreshortening effects. 
Second, the separation angle $\gamma$ is also not very large, which is expected to adversely impact the accuracy of the method: according to the tests on numerical simulations discussed in \PaperI{}, the \SDM{} accuracy at $\gamma \approxeq 27^\circ$ is close to 100\% in smooth field areas (see \eg{} Figure~10c in \PaperI{}), while it can be significantly lower in quiet-sun regions (see \eg{} Figure~10e in \PaperI{}, where \SDM{} yields the correct disambiguation in 75\% of the pixels in that test). 
We note the specific combination of $\gamma$ and spacecraft distance from the Sun presented here are not explicitly included in any of the tests presented in \PaperI{}, so the comparison between numerical tests and application to observations can only be indicative.
}

\subsection{\HRT{}}\label{s:data_hrt}
%
%
On March 17th, the \HRT{} observed the target AR for 30 minutes with a high-cadence program (the Nanoflare-Solar Orbiter Observation Plan in \cite{Zouganelis2020}). 
The dataset used in this work has Data IDentification (DID) number 0243170227 with observation time 03:44~UT. 
At the distance of 0.38\,AU, the \HRT{} pixel scale at disk center is equal to 137.2\,km.
The 24 polarization images used to build the Stokes vector were acquired with a fast accumulation mode lasting 60\,s.

The spectropolarimetric observations were calibrated and the inversion of the Radiative Transfer Equation (RTE) followed \cite{Sinjan2022}.
\kd
{In short, dark-current and flat-field corrections are first applied.
The flat-field is preliminary corrected using unsharp masking with a Gaussian amplitude of 69 pixels. 
Since the \HRT{} image stabilization system \citep{Volkmer2012} was not in operation during the acquisition, a co-registration of the polarization images is also applied, see also \cite{Calchetti2023a}.
}

\kd
{After demodulation and cross-talk correction, the RTE-inversion is performed using the CMILOS code \citep{OrozcoSuarez2007}, which assumes a Milne-Eddington atmosphere and employs analytical response functions to build the synthetic profiles that are used in the minimization process.  
The filling factor is assumed to be unity.   
A detailed description of the above steps is found in \cite{Sinjan2022}.
The above constitutes the currently  standard version of the \HRT{} pipeline,} and no additional processing was applied to the data.
In particular, the employed data have not been reconstructed and aberration-corrected, as described  by \cite{Kahil2023}, see \sect{errors} for further details. 

In addition to the disambiguation using the \SDM{}, we also performed a 
disambiguation of the same dataset using a classical method that, unlike the \SDM{}, 
only uses data from a single telescope.  
The disambiguation that we applied, hereafter \ME{},  
is an adaptation of the disambiguation code in the \HMI{} pipeline \citep{Hoeksema2014a}, which, in turn, implements the ``minimal energy method'' \citep{Metcalf1994,Metcalf2006}.  
The parameters used for the \ME{}-disambiguation are similar to those used in the HMI pipeline, and no particular attempt was made to optimize them.
However, the \ME{} method requires the definition of a noise mask to determine where to apply annealing.
This was built as a linear mask with the corresponding parameters \texttt{bthresh1}=300\,G and \texttt{bthresh2}=400\,G, which represent upper levels of noise on the transverse component of the field at disk center and at the limb, respectively. 

In order to fix \texttt{bthresh1,2}, the noise level on the \HRT{} magnetogram is estimated in two ways: first, as the standard deviation of the Gaussian fit of the histogram distribution of transverse field values in quiet Sun areas. 
Such an estimation results in a noise level on the transverse component equal to 64\,G (and of 8.47\,G on $\Blos$, see also \cite{Sinjan2023}).
Second, with a similar method as the above, the noise on the Stokes components are found  to be $(1.73,1.27,1.34)\times 10^{-3}$ for $(Q,U,V)/I_c$, respectively.
Using these values in the magnetographic formulae for classical calibration as given by Eq.~4 in \cite{Martinez-Pillet2007}, we find a  second estimate for the noise on the transverse component equal to 147\,G (and 8.89\,G on $\Blos$).
The employed values for \texttt{bthresh1} and \texttt{bthresh2} correspond to approximately three times the average of such noise estimations, for observations at center of disk and limb, respectively.

\subsection{\HMI{}}\label{s:data_hmi}
The \HMI{} magnetograms used for the \SDM{} application are chosen to match as close as possible the \HRT{} observation time corrected for the difference in light-travel time (step 1 in \sect{implementation}).  


Two \HMI{} datasets from two different \HMI{} data product series are considered.
The first dataset is the \HMI{} vector magnetogram with the observation date of 17-Mar-2022 03:46:35.5 from the \hmiME{} series (corresponding to the observation time as registered by the FITS keyword \texttt{T\_OBS}=03:47:21~UT).
\kd
{For the considered data sets, the}
difference in light travel time between \SO{} and \SDO{} is 307.6\,s.
The \hmiME{} series provides vector magnetograms (inclination, ambiguous azimuth, and field strength) at a cadence of 720\,s resulting from averages of observations taken over 20 minutes \citep{Hoeksema2014a},  
\kdtwo{and produced by the  Very Fast Inversion of the Stokes Vector
(VFISV) Milne-Eddington code \citep[][]{Borrero2011,Hoeksema2014a}.} 
The price to pay for having the vector information 
is that, due to the averaging procedure, the observation are generally more difficult to match in time with \HRT{} observations.    

The second dataset is the \HMI{} magnetogram obtained on 17-Mar-2022 03:48:28.0 from the 45 seconds data series, \hmiBlos{}.
For this dataset,  
\texttt{T\_OBS}=17-Mar-2022 03:48:51~UT.
The \hmiBlos{} data series provides the LoS magnetic field only but, thanks to its high-cadence, the \HMI{} observation time can be chosen to be very close to the time of the \HRT{} dataset.
\kdtwo{The \hmiBlos{} data series employs a ``MDI-like'' inversion method \citep{Couvidat2012} that is based on a discrete Fourier expansion of the solar neutral iron line, corrected for the \HMI{} filter transmission profile \citep{Hoeksema2014a,Couvidat2016}.
\cite{Hoeksema2014a} showed that, in areas where $|\vB|>300$\,G, the values obtained by the RTE-inversion in the \hmiME{} series are larger than those obtained by MDI-like algorithm used for the \hmiBlos{} data series (see Fig.~17 in \cite{Hoeksema2014a}). 
}

\kdtwo{
In addition to the above, we also considered a dataset from the \texttt{hmi.ME\_90s} series, which is a series similar to the \hmiME{} series but with a cadence of 90\,s instead of 720\,s.
Data products for such a series are available on demand, and were not readily available for the relevant date.
For this reason, we do not employ the \texttt{hmi.ME\_90s} series in this article, although the results from a few ancillary tests made using that series are presented in \sect{discussion}.
 }

\section{\SDM{}-disambiguation 
using the \hmiME{} series}\label{s:sdm_720_reverse}
\begin{figure}
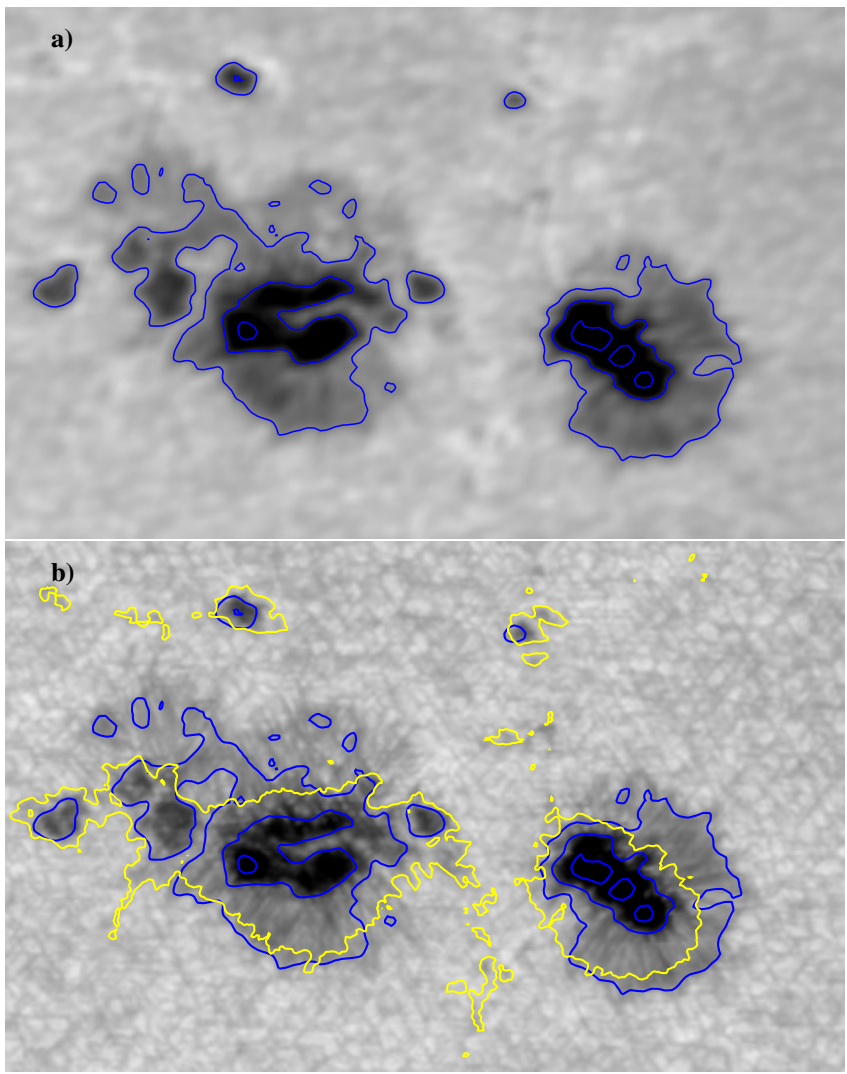

 \setlength{\imsize}{0.8\columnwidth}
 \centering
 \labfigure{\imsize}{\figME/fig2a}{0.03\imsize}{0.59\imsize}{a)}\\
 \labfigure{\imsize}{\figME/fig2b}{0.03\imsize}{0.59\imsize}{b)}
 \\
 \caption{Continuum images of a) remapped-\HMI{} and b) \HRT{}
  \kd{, on the \HRT{} image plane}. 
  \kd{The continuum intensity is normalized by its median value, and shown between 0.5 and 1.2.  
  On both panels, the isocountours of the \HMI{} continuum intensity at $[15.,25.,35.]\times 10^3$\,DN\,s$^{-1}$ (corresponding to $[0.37, 0.62,0.86]$ in normalized units) are drawn as blue solid lines.
  In panel b, the 400 G-isoline of the \HRT{} $|\Blos|$ is drawn as a yellow solid line.
  }
}
\label{fig:cnt_720_vect}
\end{figure}
%
%
In this section we present the results of the disambiguation of the \HRT{} vector magnetogram using the \SDM{} and the \hmiME{} data series as input 
\kdtwo{(\ie{} the ``reverse'' case, with A=\HRT{} and B=\HMI{}, see \sect{application})}.
First, steps 2 and 3 of the procedure in \sect{implementation} are performed, using the field strength as co-registration field.
The subdomain used for co-registration (and \SDM{} application) is the blue rectangle on the right panel of \fig{fov}. 
As a result, the \HMI{} magnetogram is co-registered with, re-projected onto, and remapped on the \HRT{} image plane, \kd{to which we refer hereafter} 
as the ``remapped-\HMI{}'' magnetogram.
For the nominal helioprojective coordinates of the reference pixel \texttt{CRVAL1,2}=(416.36,1281.19), the co-registered reference pixel values are found to be \texttt{CRPIX1,2}=(1230.72, 983.07). 
The images co-registration is then the same for the two applications in \sect{sdm_720_reverse_tr} and \sect{sdm_720_reverse_los}. 
\kd{A qualitative check of the co-registration procedure obtained using the field strength  is given by} \Fig{cnt_720_vect}a,b, which show the continuum intensity of the remapped \HMI{} and \HRT{}, respectively, with the isolines of the \HMI{} continuum overlaid on both images. 
The co-registration of the two datasets is \kd{globally} very accurate, \kd{with a (Pearson) correlation coefficient of the field strength equal to 0.97.}

\begin{figure*}
 \setlength{\imsize}{\textwidth}
 \centering
 \labfigure{0.48\imsize}{\figME/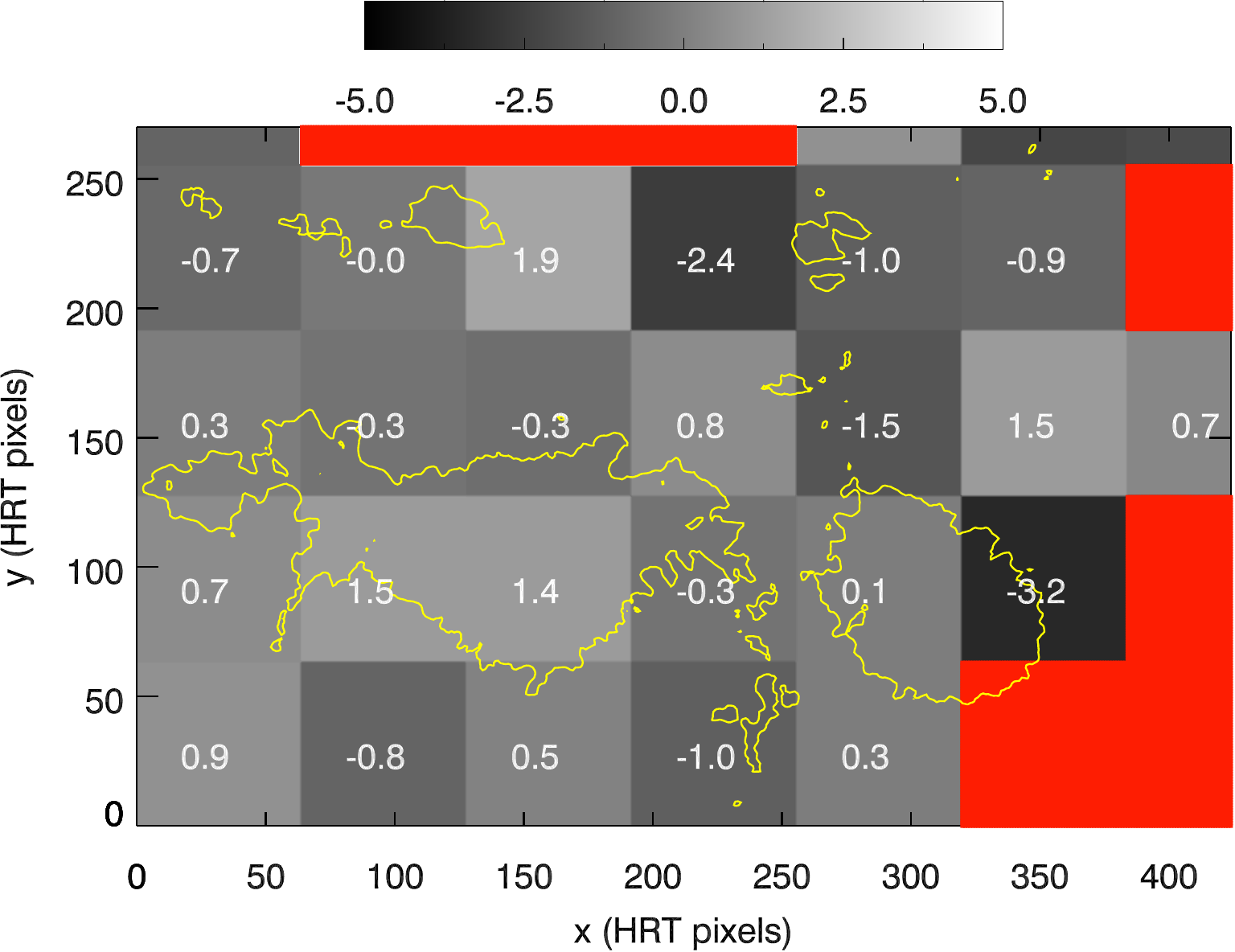}{0.03\imsize}{0.35\imsize}{a)}
 \labfigure{0.48\imsize}{\figME/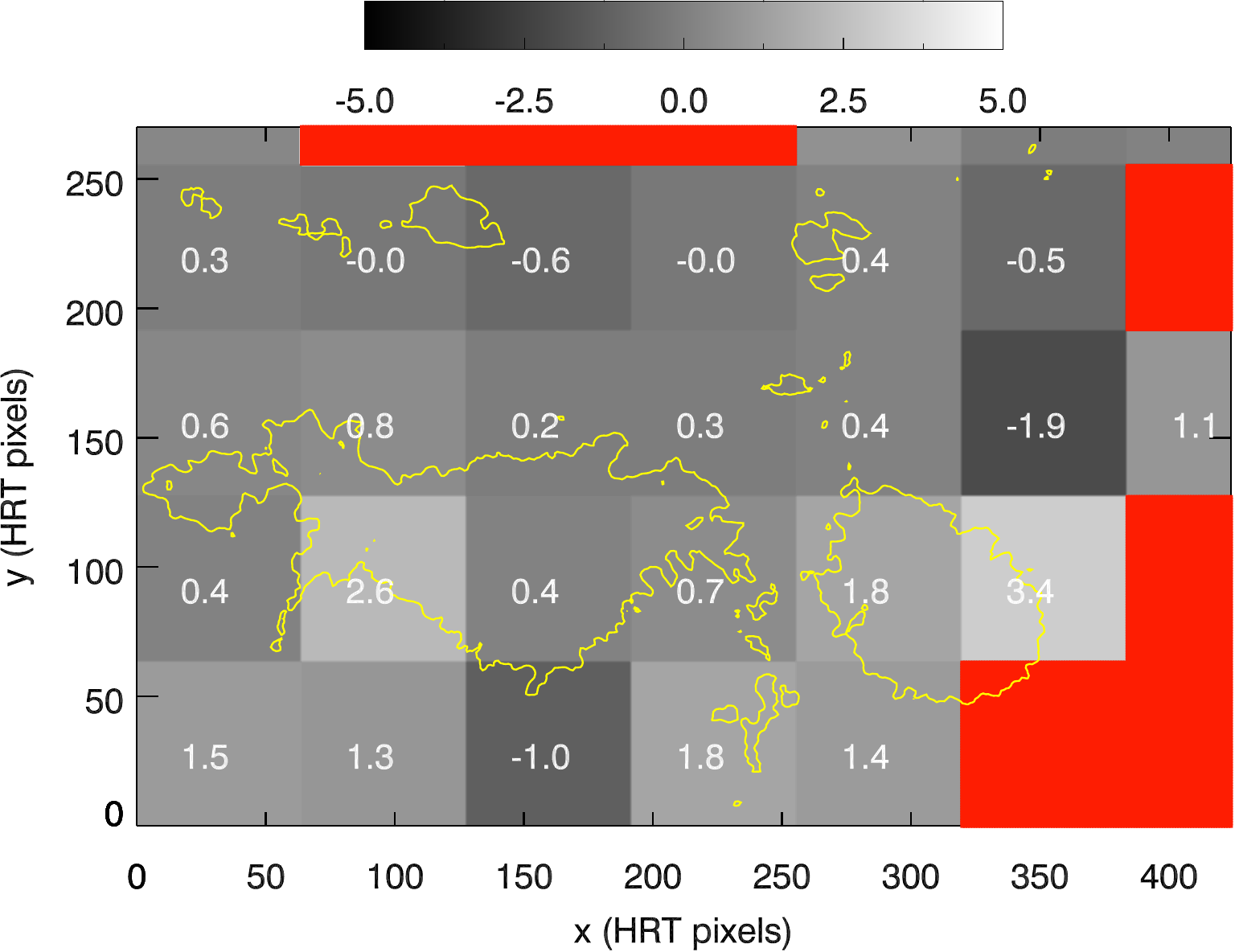}{0.03\imsize}{0.35\imsize}{b)}
 \\
 \caption{
  \kd{Residual cross-correlation shifts after co-registration, computed in tiles of 64$\times$64 pixels, in $x$- (panel a) and $y$-direction (panel b). 
The number at the center of each tile is the co-registration shift for that tile, in units of \HRT{} pixels. 
Tiles where the cross-correlation procedure did not converge are marked in red. 
In both panels, the 400 G-isoline of the \HRT{} $|\Blos|$ is drawn as a yellow solid line.
}}
\label{fig:cnt_720_cc}
\end{figure*}
\kd{In order to quantitatively further describe the co-registration accuracy, we divide the co-registration subdomain in tiles of 64$\times$64 pixels, and recompute the co-registration shifts between the remapped-\HMI{} and  the (co-registered) \HRT{} images in each tile separately, following the same iterative procedure as step 2 in \sect{implementation}.
This results in the maps of residual misalignment shown in \fig{cnt_720_cc}a,b, which shows that, with few exceptions, all residual shifts are within $\pm2$ pixels.
The residual shifts are, however, often of variable magnitude and different sign, even between adjacent tiles. 
As discussed more extensively in \sect{errors}, this can likely be attributed to uncorrected aberrations in the employed \HRT{} dataset.   
}

Next, either \eq{sdm_tr} or \eq{sdm_los} is applied to remove the ambiguity of the \HRT{} transverse field. 
\subsection{First observation-based disambiguation}\label{s:sdm_720_reverse_tr}
\begin{figure*}
 \setlength{\imsize}{0.80\textwidth}
 \begin{center}
 \labfigure {\imsize} {\figME/fig4a} {0.03\imsize} {0.70\imsize}{a)}
 \vspace{5pt} \\
 \labfigure {\imsize} {\figME/fig4b}{0.03\imsize}{0.70\imsize}{b)}
 \end{center}
 \caption{First observation-based \SDM{}-disambiguated vector magnetogram. 
   The \HRT{} vector magnetogram is disambiguated using \HMI{} information from the (ambiguous) \hmiME{} data series and \eq{sdm_tr}.
  \kd{Panel a:} The background image represents $\Blos$ \kd{on the \HRT{} image plane,} saturated at $\pm1500$\,G in greyscale, and red/blue arrows represent the transverse field at positive/negative $\Blos$, respectively.
 The 400\,G-isoline of $|\Blos|$ is drawn as a \kd{yellow} solid line. 
 \kd{Panel b)  same as panel a but in remapped Stonyhurst coordinates, with the magnetic field reprojected in radial, poloidal and toroidal components. 
 In this panel, the background image  represents the radial component $B_r$ saturated at $\pm1500$\,G in greyscale, and red/blue arrows represent the horizontal field at positive/negative $B_r$, respectively.
 The 400\,G-isoline of $|B_r|$ is drawn as a yellow solid line.} 
}
\label{fig:reverse_720_vect_mgm}
\end{figure*}
We first consider the application of \eq{sdm_tr}.
Following steps 3 to 7 in \sect{implementation}, a parity map is produced that takes in each \HRT{} pixel the value $-1$ (respectively, $+1$) where the transverse component is (respectively, is not) to be reversed. 
Applying the parity map to the \HRT{} ambiguous transverse component we obtain the first \SDM{}-disambiguated vector magnetogram, shown in \fig{reverse_720_vect_mgm}\kd{a}. 
On a qualitative level, the \SDM{}-disambiguation of the \HRT{} magnetogram is remarkably successful.
In  particular, the transverse component has the expected orientation, namely it is pointing radially outward in positive flux concentrations. 
The transverse component is also smoothly distributed almost everywhere on the main polarities. 
This is true also for the bottom part of the AR, where projection effects shorten the amplitude of the transverse component considerably. 
Such properties are remarkable if one recalls that, in order to produce the \SDM{}-disambiguated transverse field, no assumption about the transverse field is made: the disambiguation in \fig{reverse_720_vect_mgm} solely results from combining information from two points of view (two telescopes) in each \HRT{}-pixel separately.  
Localized areas where the orientation of the transverse component is different from that expected are discussed in \sect{diag}, whereas general accuracy considerations are summarized in \sect{discussion}.

\subsection{Disambiguation diagnostic}\label{s:diag}
\begin{figure*}
 \setlength{\imsize}{0.8\columnwidth}
 \centering
 \labfigure{\imsize}{\figME/fig5a}{0.03\imsize}{0.72\imsize}{a)}
 \labfigure{\imsize}{\figME/fig5b}{0.03\imsize}{0.72\imsize}{b)}
 \caption{Application of the \SDM{} using the \hmiME{} series and \eq{sdm_tr}.
  Panel a shows the sign function $\sgn$, saturated at $\pm5$;
  panel b shows the corresponding \SDM{}-disambiguated vector magnetogram, with $\Blos$ saturated at $\pm1500$\,G in greyscale and red/blue arrows representing  the transverse field at positive/negative $\Blos$ (same as \fig{reverse_720_vect_mgm}a).
  The rectangles in panel b indicate the areas of suspicious disambiguations discussed in \sect{diag} (\kd{black}) and \sect{diag_sgn} (orange and green).   
  \kd{In both panels, the 400\,G-isoline of the \HRT{} $|\Blos|$ is drawn as a yellow} solid line. 
 }
\label{fig:reverse_720_vect_tr} 
\end{figure*}
Besides the overall consistency of the disambiguation in \fig{reverse_720_vect_mgm}, there are specific locations where the orientation of the transverse component is suspicious.
Two such areas follow the \kdtwo{polarity inversion line  of $\Blos$ (\PIL{})} in the penumbral areas, highlighted by the orange rectangles at $(x,y)\simeq(180,170)$ and $(x,y)\simeq(340,130)$ in \fig{reverse_720_vect_tr}b. 
While in most places the blue arrows continue pointing away from the spots, as expected in the crossing of the \kdtwo{\PIL{}} in the penumbra (\ie{} where arrows change color from red to blue), at these locations, the blue arrows point in the opposite direction to the red arrows and are seemingly interlaced with them.  

Another two areas of suspicious disambiguation lie along linear features at $(x,y)\simeq(180,110)$ and $(x,y)\simeq(330,\kd{80})$
(highlighted by \kd{black} rectangles in \fig{reverse_720_vect_tr}b), where the arrows representing the transverse field are left-oriented, opposite in orientation to those right above and below that line but of similar amplitude, and, therefore, arguably pointing in the wrong direction.  

The real orientation of the transverse component in \fig{reverse_720_vect_mgm} is not known, and a quantitative assessment of the \SDM{} accuracy (as well as of any other method) is not possible in such a case.
However, and \kd{in contrast to} 
traditional single-view methods, the \SDM{} offers diagnostic tools that help assess at which specific locations the method is likely to be less accurate.

\kd
{The evaluation of the correctness of the disambiguation is customarily performed writing the field in the ``heliographic'' \citep{Gary1990} radial, toroidal, and poloidal components \citep[or, more precisely, in heliocentric spherical coordinates, see][]{Thompson2006,Sun2013}. 
In addition, in order to compensate for foreshortening effects, the field is remapped (\ie{} interpolated) onto the Stonyhurst coordinates \citep{Thompson2006}. 
The resulting vector magnetogram in heliographic projection is shown in  \fig{reverse_720_vect_mgm}b, where the radial field is represented by the greyscale background with a 400\,G isoline in yellow, and the horizontal field is represented by blue/red arrows in correspondence of negative/positive values of the radial field, respectively.
The polarity inversion line, not drawn, is the line separating blue from red arrows.
One can clearly see how some of the suspicious areas discussed above affect the re-projected field shown in \fig{reverse_720_vect_mgm}b.
In particular, the black boxes in  \fig{reverse_720_vect_tr}b correspond to areas (slightly shifted downwards by the re-projection) where the radial field is smaller than in their surroundings, yielding darker (but still of positive value) structures in the cores of the spots.
Even more remarkably, a true reversal of the radial component can be seen in correspondence of the penumbral area highlighted by the right-orange box in \fig{reverse_720_vect_tr}b: here the horizontal field is drawn in blue arrows, meaning that corresponds to negative values of the radial component.
Such effects are a clear example of how errors in the disambiguation of the transverse field component can heavily affect all (physically-relevant) field components in the local frame. 
On the other hand, the areas marked by the left-orange and green box in \fig{reverse_720_vect_tr}b are not readily recognizable as suspicious in \fig{reverse_720_vect_mgm}b.}

\kd
{In our identification of suspicious areas, the assumption that the transverse component is smooth across neighbouring pixels is implicitly made, which has some similarities with the hypothesis that underlies the \ME{} method.
}
\kd
{
However, such an assumption does not enter at any point the derivations of \eqs{sdm_tr}{sdm_los}. 
Moreover, in the following we introduce diagnostic quantities that are able to identify exactly such areas as locations of potentially wrong disambiguations regardless of any continuity assumption.
Since such diagnostic quantities are defined in the image plane rather than in the heliographic plane, we prefer to discuss them primarily in the former rather than remap them onto the latter, thereby avoiding the loss of the pixel-by-pixel connection to the observed quantities appearing in \eqs{sdm_tr}{sdm_los}.
}

\subsubsection{The sign function $\sgn$}\label{s:diag_sgn}
In principle, the sign function $\sgn$ in \eq{sdm_tr} (as well as in \eq{sdm_los}) should take only the values $\pm1$.
However, since $\sgn$ is determined as a combination of field components from different instruments, fluctuations due to noise or any difference in observation time, instrument calibration, RTE inversion, or co-registration would produce departures from the nominal values. 
Conversely, the departure of \eq{sdm_tr} from the nominal  $\pm1$ values can be used as a diagnostic metric for the \SDM{}.

The application of \eq{sdm_tr} results in the map of $\sgn$ that is shown in \fig{reverse_720_vect_tr}a, where large departures from the nominal $\pm1$ values show up as black/white pixels, mostly, but not exclusively, in low-field\kd{, quiet-sun areas outside the 400\,G yellow isoline}. 
Discarding the latter, the most prominent area of errors is around $(x,y)\simeq(280,\kd{80})$, marked by the green rectangle in \fig{reverse_720_vect_tr}b, which was recognized to be an area of low-signal in the \kd{\HRT{}} linear polarization yielding difficulties in the inversion. 

Other locations where the sign function $\sgn$ significantly departs from the nominal values is the \kdtwo{\PIL{}} of $\Blos$, indeed where the orange rectangles in \fig{reverse_720_vect_tr}b indicates suspicious disambiguations.
At this \kdtwo{\PIL{}}, the transverse field is relatively large. 
However, the LoS component is not, and may fluctuate due to noise, or difficulties in the inversion from atypical Stokes~V profiles \citep[see][]{Solanki1993}.
Since $\Blos$ is the only term at the numerator of \eq{sdm_tr}, such fluctuations can easily be the origin of the ``salt and pepper'', large values of $\sgn$ found in these areas. 

In addition, we notice the roughly horizontal separation line between $-1$ (bottom) and $+1$ (top) values across the main polarity concentrations (before disambiguation, all arrows point upwards because of $\alA$ being restricted to $[0;\pi]$).
Such a line is not a ``special'' place in any physical sense: it is ultimately determined by the mutual orientation and locations of the spacecraft (via $\thA$ and $\thB$) and the particular distribution of the observed field (via $\alA$).
However, the horizontal transition line between $\sgn=+1$ and $\sgn=-1$ is spatially correlated to the linear distribution of suspicious arrows noticed above.

\subsubsection{The $\Bn$-components}\label{s:diag_Bn}
\begin{figure}
 \setlength{\imsize}{0.8\columnwidth}
 \centering
 \labfigure{\imsize}{\figME/fig6a}{0.03\imsize}{0.72\imsize}{a)}\\
\labfigure{0.9\imsize}{\figME/fig6b}{-0.01\imsize}{0.78\imsize}{b)}
 \caption{Relative difference $\dBn$ \kd{on the \HRT{} image plane}, see \eq{Bn}.
 a) spatial distribution of $\delta \Bn$; the 400\,G-isoline of the \HRT{} $|\Blos|$ is drawn as a solid, black  line; the purple slit at x=170 corresponds to the location of the one-dimensional plot in panel b.  
 b) Upper panel: profiles of the $\Bn$ components along the purple slit in panel a, for the remapped-\HMI{} (green) and \HRT{} (purple); bottom panel: corresponding $\dBn$ along the purple slit in panel a.
} 
\label{fig:reverse_720_Bn} 
\end{figure}
In order to have a more quantitative analysis of the above disambiguation result we exploit the property of the \SDM{} that the $\Bn$-component of \HRT{} and the $\Bn$-component of the remapped-\HMI{} should be identical (see \sect{geometry}). 
Such a property should be considered a pre-requisite for application of the \SDM{}, and pixels where the property is not fulfilled are expected to be more prone to disambiguation errors. 
On the other hand, it is necessary condition for the application of the \SDM{} that involves only the transverse components, meaning that disambiguation errors that are originated from the LoS components only can still occur even if the two $\Bn$ component are identical. 
Hence, we can use the metric 
\BE
\dBn=\frac{\Bn^{\rm PHI}-\Bn^{\rm HMI}}{\Bn^{\rm PHI}+\Bn^{\rm HMI}}
\label{eq:Bn}
\EE
as a measure of how well the assumptions of the \SDM{} are fulfilled in the practical application.
Since $\Bn$ is by construction positive, 
$\dBn$ varies between $-1$ and $+1$, and takes the value zero where the two components are identical. 

We notice that a metric similar to \eq{Bn} can be equally constructed using the field strength $|\vB|$, which is also theoretically independent of the point of view.
Tests not reported here show, however, that no information that is relevant to the \SDM{} result is really gained in this way: 
differences in field strength between \HRT{} and \HMI{} are relatively small, as documented by \cite{Sinjan2023}, and do not discriminate the role of the azimuth in any way.

The relative difference $\dBn$ is shown in \fig{reverse_720_Bn}a.
Except for purely quiet-sun areas, $\dBn$ is indeed found to be close to zero everywhere, which indicates that most of the considered FoV fulfills the requirements for the application of the \SDM{}.
However, a remarkable double-stripe of opposite $\dBn$ unitary values is evident around y$\simeq 100$.
This double-stripe structure in $\dBn$ is due to an apparent shift in the vertical direction of the two $\Bn$ distributions.
The apparent shift is confirmed by the plot of the $\Bn$-components in the upper panel of \fig{reverse_720_Bn}b, which is taken along the purple slit in \fig{reverse_720_Bn}a.
The shift is about 8 \HRT{}-pixels in \fig{reverse_720_Bn}b, and varies between 5 and 9 pixels, depending where the slit is placed on the double-stripe structure.
The \HMI{}-$\Bn$ component (in green in the upper panel) is smoother than the \HRT{} one because of the averaging procedure employed in the production of the dataset, and because the \HMI{} image is upscaled when remapped to the \HRT{} detector plane 
(\HRT{} resolution being 2.6 times higher than \HMI{} in the given spacecraft configuration). 

Additional minor differences can be seen in isolated pixels of \Fig{reverse_720_Bn}a (including the area around (\kd{280},80) noticed already above), and may be due to differences in calibration and RTE inversion (see discussion in \sect{errors}).
Similarly, isolines close to unity of one-pixel width connecting the left sunspot to the upper area are visible in the area around (100,180).
These correspond to locally vanishing values of the remapped-\HMI{} $\Bn$ component, and should be labelled as locations of potential inaccuracy of the disambiguation in specific pixels.
However, besides such smaller differences, a clear and more significant shift in the location of $\Bn=0$ between the two maps is incontrovertible, which shows that the assumption for a meaningful application of the \SDM{} are violated in the double-stripe area.

In order to understand the consequences of the $\dBn$ double-stripe on the accuracy of the \SDM{} let us recall the employed polar representation of the transverse component from \sect{geometry}. 
First, $\Bn$ is defined by \eq{components} as a sine-function, and does not enter either of \eqs{sdm_tr}{sdm_los}. 
On the other hand, the $\Bw$ component enters both equations, and, since it is defined as a cosine function, it has large jumps at the same locations where $\Bn=0$, \ie{} at the ends of the polar angle interval of definition $[0,\pi]$.
Since the difference between the $\Bw$ components estimated from each telescope enters the denominator of \eq{sdm_tr}, errors in the $\Bw$-components close to location where $\Bn=0$ are amplified by the discontinuous character of the $\Bw$-components there, thereby heavily affecting the \SDM{} accuracy at such locations.
Indeed, the double-stripe exactly corresponds to the locations of suspicious disambiguations marked by the \kd{black} rectangles in \fig{reverse_720_vect_tr}b.
A discussion of the possible origin of the double-stripe is presented in \sect{errors}. 

\begin{figure}
 \setlength{\imsize}{0.8\columnwidth}
 \centering
 \labfigurecr{\imsize}{\figME/fig7a}{-0.03\imsize}{0.55\imsize}{a)}{0}{90}{0}{0}
 \\
 \labfigurecr{\imsize}{\figME/fig7b}{0.0\imsize}{0.55\imsize}{b)}{0}{90}{0}{0}
 \\
 \labfigure{\imsize}{\figME/fig7c}{-0.03\imsize}{0.60\imsize}{c)}
 \caption{
 Comparison of \SDM{}- and \ME{}-disambiguations \kd{on the \HRT{} image plane.} 
 The \SDM{} is applied using information from the \hmiME{} series and:
 \kd{Panel a}: \eq{sdm_tr}, see \sect{sdm_720_reverse_tr}.
 \kd{Panel b}: \eq{sdm_los}, see \sect{sdm_720_reverse_los}.
 \kd{Panel c}: a combination of  both \eq{sdm_tr} and \eq{sdm_los}, see \sect{best}.
 Agreement is indicated in \kd{black}, 
 disagreement in \kd{white}, 
 \kd{whereas grey} indicate pixels that are not considered in the comparison because the \HRT{} transverse field falls below the noise threshold; the 400\,G-isoline of \kd{the \HRT{}} $|\Blos|$ is drawn as a \kd{yellow} solid line; axes units are \HRT{} pixels.
} 
\label{fig:reverse_720_ME0} 
\end{figure}
\subsection{Comparison with \ME{}}\label{s:me0}
Next, we compare the \SDM{}-disambiguation with the standard, single-view \ME{}-disambiguation obtained as described in \sect{data_hrt}.
While this is not a test for the \SDM{} for the reason explained in \sect{data_hrt}, it is useful to show where differences are located that deserves further investigation.

In the comparison between the two disambiguation methods we exclude pixels where the \HRT{} transverse field is not annealed by the \ME{} method, or where the transverse component is below 400\,G (\ie{} \kd{below} the threshold set by the noise on the transverse component, see \sect{data_hrt}). 
The rate of agreement of \SDM{} and \ME{} disambiguations in this domain is 84.4\%.
\Figs{reverse_720_ME0}a render visually the spatial distribution of this agreement, where in \kd{black} are represented pixels where the two methods agree, in \kd{white} where they do not, and in \kd{grey} pixels that are excluded from the comparison according to the criteria above.

The \SDM{}- and \ME{}-disambiguations agree in most of the analysed domain, with two exceptions: 
first, on the location of the double-stripe structure of \fig{reverse_720_Bn}a and, second, along the \kdtwo{\PIL{}} in the northern penumbral areas.
The former disagreement area is expected from the violation of the $\dBn=0$ necessary condition. 
The latter disagreement is also expected because the sign function $\sgn$ has values largely departing from the nominal ones in that area (\fig{reverse_720_Bn}a).
We notice that, according to our interpretation in \sect{diag_sgn}, these errors are due to fluctuations of the sign of $\Blos$ close to its \kdtwo{\PIL{}}, which occur even though $\dBn=0$ there. 
Hence, both disagreements are expected from the discussion above, and they are likely true but remediable errors of the \SDM{} (see \sect{discussion} and \sect{best} in particular). 
\subsection{Comparison of disambiguations using \eq{sdm_tr} and \eq{sdm_los}}\label{s:sdm_720_reverse_los}
\begin{figure*}
 \setlength{\imsize}{0.8\columnwidth}
 \centering
 \labfigure{\imsize}{\figME/fig8a}{0.03\imsize}{0.72\imsize}{a)}
 \labfigure{\imsize}{\figME/fig8b}{0.03\imsize}{0.72\imsize}{b)}
 \caption{\SDM{}-disambiguated vector magnetograms using the \hmiME{} series and \eq{sdm_los}.
  Panel a shows the sign function $\sgn$, saturated at $\pm5$;
  panel b shows the vector magnetogram
  \kd{on the \HRT{} image plane}, 
  with $\Blos$ saturated at $\pm1500$\,G in greyscale and red/blue arrows representing  the transverse field at positive/negative $\Blos$.
  \kd{In both panels, the} 400\,G-isoline of \kd{the \HRT{}} $|\Blos|$ is drawn as a \kd{yellow} solid line.
\kd{The corresponding vector magnetogram in heliographic projection is shown in \fig{app_helio}a.}
 }
\label{fig:reverse_720_vect_los} 
\end{figure*}
In this section we consider the \SDM{}-disambiguation that is obtained by applying \eq{sdm_los} instead of \eq{sdm_tr} to the same data and procedure as in \sect{sdm_720_reverse_tr}. 
\kd{In this case, from the \hmiME{} series, only the LoS information is used in \eq{sdm_los}. 
On the other hand,} since the input data are the same, the co-registration and remapping procedure is identical to the one in \sect{sdm_720_reverse_tr}.

The sign function $\sgn$ obtained by the application of \eq{sdm_los} is shown in \fig{reverse_720_vect_los}a, with the corresponding disambiguated magnetogram in \fig{reverse_720_vect_los}b.
\Fig{reverse_720_vect_los}a shows that, in comparison with the result from \eq{sdm_tr} in \fig{reverse_720_vect_tr}a, there are larger areas where $\sgn$ is very different from the nominal value, $\pm1$, with corresponding larger areas of obviously wrong disambiguation (see \fig{reverse_720_vect_los}b), 
\kd
{in particular on a relatively large patch around $(x,y)\simeq(115,115)$ and, to a smaller extent, around $(x,y)\simeq(290,110)$.
The incorrect disambiguation on such areas determines a reversal of the radial component when the vector magnetogram is reprojected in heliographic coordinates, see \fig{app_helio}a.
} 

The mechanism by which errors in \eq{sdm_los} occur is analogous to \eq{sdm_tr} but involves the difference between the LoS-components in the numerator of \eq{sdm_los}.
Arguably, results are worse for \eq{sdm_los} than for \eq{sdm_tr} because the angle $\gamma$ is relatively small and differences between the two $\Blos$ are more affected by errors, but see also \sect{errors} for further discussion.

The comparison with the \ME{} disambiguation is shown in \fig{reverse_720_ME0}c,d. 
The agreement between the two methods in this case is lower than for \eq{sdm_tr}, being equal to 71\%.
The lower level of agreement is understood as the result of the larger areas where $\sgn$ is more strongly departing from the nominal values $\pm1$ with respect to the case in \fig{reverse_720_vect_tr}a.

On the other hand, some of the penumbral areas to the north and west of the sunspots (at the location of the orange rectangles in \fig{reverse_720_vect_tr}b) are consistently agreeing with \ME{} results, differently from what happens for \eq{sdm_tr} (cf. the two top \kd{panels} in \fig{reverse_720_ME0}).
Indeed, the sign obtained by \eq{sdm_los} in these areas is homogeneously close to the nominal value, yielding a consistent disambiguation also along the \kdtwo{\PIL{}} of the (\HRT{}) $\Blos$.
The reason for this improvement is that \eq{sdm_los} involves a difference between the two $\Blos$, which, due to the different views, have \kdtwo{\PIL{}} at different locations, and is therefore less affected by fluctuations of the individual LoS components at the correspondent \kdtwo{\PIL{}}.  
We return to this complementarity of results in \sect{best}.   

\section{\SDM{}-disambiguation using the \hmiBlos{} series}\label{s:sdm_45_reverse_los}
\begin{figure}
 \setlength{\imsize}{0.7\columnwidth}
 \centering
 \labfigure{\imsize}{\figBlos/fig9a}{0.03\imsize}{0.72\imsize}{a)}
\\
 \labfigure{\imsize}{\figBlos/fig9b}{0.03\imsize}{0.72\imsize}{b)}
\\
 \labfigure{\imsize}{\figBlos/fig9c}{0.03\imsize}{0.65\imsize}{c)}
 \caption{
 \SDM{}-disambiguated vector magnetogram using information from the \hmiBlos{} series, see \sect{sdm_45_reverse_los}, \kd{on the \HRT{} image plane}.
 \kd{Panel a):} sign function $\sgn$ from \eq{sdm_tr}, saturated at $\pm5$;
 \kd{panel b):} vector magnetogram, with $\Blos$ saturated at $\pm1500$\,G in greyscale and red/blue arrows representing the transverse field at positive/negative $\Blos$;
 \kd{panel c): comparison of \SDM{}- and \ME{}-disambiguations, with agreement indicated in black, disagreement in white, and grey indicates pixels that are not considered in the comparison because the \HRT{} transverse field falls below the noise threshold.
 In all panels, the 400\,G-isoline of \kd{the \HRT{}} $|\Blos|$ is drawn as a yellow solid line.
}
\kd{The corresponding vector magnetogram in heliographic projection is shown in \fig{app_helio}b.} 
}
\label{fig:reverse_blos_vect}
\end{figure}
In this section we use the \hmiBlos{} series as input to the \SDM{}\kd{, instead of the \hmiME{} series as in \sect{sdm_720_reverse}}. 
The \hmiBlos{} data series provides the LoS magnetic field only, at a cadence of 45\,s. 
\kd{In principle, given the shorter integration time of the \hmiBlos{} data series with respect to the \hmiME{} data series, the latter  is more similar to the \HRT{} observation, which has 60\,s integration time.
This speculation is indeed confirmed by the higher correlation coefficient between the LoS components of  \HRT{} and \HMI{} found by \cite{Sinjan2023}, see in particular their Table~3.
The question is then if such an increased temporal homogeneity between \HRT{} and \HMI{} inputs to the \SDM{} is reflected in an increased accuracy of the resulting disambiguation.}

\kd
{On the other hand, changing to the \hmiBlos{} data series has also some unavoidable consequences.
First, while the availability of $\Blos$ only is not an obstacle for the application of the \SDM{}, in such a case the co-registration procedure (step 2 in \sect{implementation}) is in principle less accurate, especially at large separation angles $\gamma$, because the different viewing angles lead to different $\Blos$. 
Such an effect can indeed be a serious limitation to application of \SDM{} when one telescope provides only LoS information and the separation angle $\gamma$ is large. 
An accurate co-registration must be provided in other ways in such cases.
Alternatively, the continuum intensity can be used instead of the magnetic field in the co-registration procedure.
However, taking the comparison of \fig{cnt_720_vect}a and b as an example, we can clearly see how the continuum intensity is also affected by both projection effects, even at relatively small separation angles,  and by the difference in the resolution of the telescopes.
For these reasons,  we prefer to use the magnetic field for the co-registration procedure in this work, when possible. 
Indeed, this very topic, namely the center-to-limb variation of continuum intensity observations, is now studied using the multi-view opportunity offered by \SO{}, see for instance \cite{Albert2023} and references therein.}

\kd{Second, since only the LoS component is available in the \hmiBlos{} series, then \eq{sdm_los} is the only equation that can be used, given that \eq{sdm_tr} requires information about the transverse component of the field obtained from both telescopes. 
Therefore, insofar as the \SDM{} equation is concerned,  the case presented in this section is similar to the one discussed in \sect{sdm_720_reverse_los}, where the same \eq{sdm_los} was applied using the LoS information from the \hmiME{} data series.}

The employed \HMI{} input to the \SDM{} procedure in \sect{implementation} is the LoS-magnetogram with the observation date 17-Mar-2022 03:48:28 from the \hmiBlos{} series (corresponding to \texttt{T\_OBS}=03:48:51~UT).
The co-registration procedure using $\Blos$ associates the nominal \texttt{CRVAL1,2} to \texttt{CRPIX1,2}=(1232.21, 983.04), which, thanks to the relative small $\gamma$, are comparable to the co-registration shifts of \sect{sdm_720_reverse}.

\Fig{reverse_blos_vect} summarizes the application of \eq{sdm_los} to this case. 
\kd{Globally, the} sign $\sgn$ in \fig{reverse_blos_vect}a has a very similar spatial distribution as in \fig{reverse_720_vect_los}a, but shows weaker departures from the nominal $\pm1$ values. 
The same is true for the disambiguated vector magnetogam (\fig{reverse_blos_vect}b) that has obvious errors in similar locations as in  \fig{reverse_720_vect_los}b.  
\kd
{More in details, the areas of departures from the nominal values of the sign function $\sgn$ (in particular around $(x,y)\simeq(115,115)$ and $(x,y)\simeq(290,110)$) appear to be  slightly smaller in \fig{reverse_blos_vect}a than in \fig{reverse_720_vect_los}a.
Correspondingly, also the areas where a reversal of the radial component is present as a consequence of disambiguation errors are smaller, as can be seen by comparing \fig{app_helio}b to \fig{app_helio}a.}
\kdtwo{Some of these differences can arguably be ascribed to the different inversion methods that the \hmiBlos{} and \hmiME{} data series employ \citep[see \sect{data_hmi} and also][for the cross-calibration of the LoS field component between \HRT{} and the two \HMI{} data series]{Sinjan2023}.}

\kd{Finally, from the point of view of the analysis of the results, since the transverse component of the \HMI{} observation is not available in this case, then no diagnostic of the transverse component $\Bn$ is of course possible, in contrast to what is done in \sect{diag_Bn}.}
The comparison with \ME{} gives a marginally better agreement (in 73\% of the selected domain) than in \sect{sdm_720_reverse_los}, with an analogous spatial distribution of the agreement/disagreement areas (cf. \fig{reverse_blos_vect}c,d with the corresponding \fig{reverse_720_ME0}c,d).  

In summary, using a higher-cadence $\Blos$-magnetogram instead of a lower-cadence/longer-averaged one does improve marginally the accuracy of the \SDM{} ($\sgn$ closer to the nominal values), but the spatial distribution of $\sgn$ is essentially the same.    

\section{Sources of error and future improvements}\label{s:discussion}

Sections~\ref{s:sdm_720_reverse} and \ref{s:sdm_45_reverse_los} describe the first applications to observed data of the \SDM{} that \PaperI{} presented and validated using numerical simulations.
Broadly speaking, the \SDM{} produces largely spatially-homogeneous and consistent disambiguations.
The comparison with the \ME{} method shows an agreement between \SDM{} and \ME{} that ranges from  84\% (\sect{me0}) to 71\% (\sect{sdm_720_reverse_los}) of the pixels in the considered subdomain of the FoV.

Single-viewpoint methods, such as \ME{}, \kdtwo{are routinely applied} (\eg{} in the \HMI{}-pipeline and in the in-development \PHI{}-pipeline), and are not restricted to the stereoscopic range of applicability as the \SDM{} is. 
On the other hand, single-viewpoint methods are invariably based on assumptions, and a validation that is based on observed data only is highly desirable. 
We show here that the \SDM{}, besides its direct application,  may become such a reliable verification tool of single-view disambiguation methods. 
Our long-term goal is to attain an accuracy that allows for such verifications.

One advantage of the \SDM{} is that it provides consistency tests to assess the quality of the obtained disambiguation in each pixel independently: first, the computed sign function $\sgn$, which should nominally be equal to either $+1$ or to $-1$, whereas \eqs{sdm_tr}{sdm_los} provide a continuous value that depend on observed quantities; second, the difference between the $\Bn$ components of the two telescopes, $\dBn$, which should be zero for perfectly compatible observations. 
In the previous sections we positively associate anomalous values of such quantities to apparent possible errors in the \SDM{}-disambiguations.
Here we discuss how such errors may arise, which tests have been made to identify their sources,  and which strategies can be taken to remove them. 

\subsection{
\kd{Discussion on possible sources of error}
}\label{s:errors}
First of all, the \SDM{} relies on a direct comparison of quantities observed by different instruments. 
The values that are compared are the product of the combined processes of spectropolarimetric data calibration and RTE inversion.
Despite a relative similarity between \HRT{} and \HMI{}, \eg{} in terms of \kd{observed spectral} line and spectral sampling \citep[see \eg][]{Solanki2020}, the two telescopes have different resolutions, which affect the \kd{retrieved} magnetic field \kdtwo{\citep[see, \eg{}][]{Stenflo1985,Leka2012,Pietarila2013,deRosa2015}}. 
Hence, a \kd{precise} cross-calibration must eventually be included in the data preparation, and progress in this direction is underway \citep[see][]{Sinjan2023}.
This should also include the effect of stray light \kd{and filling factor \citep[see, \eg{}][]{Liu2022,Leka2022}}, which might be different for \HRT{} and \HMI{}.
The effect of stray light is particularly important in sunspots, where it may lead to underestimations of the magnetic field strength \citep{LaBonte2004}.
In addition, the RTE inversion of the \HRT{} dataset employed here \citep{Sinjan2022} produces field maps of very high quality \kdtwo{with a low level of noise}. 
However, further improvements may still be possible. 
For instance, an RTE inversion \kd{of the \HRT{} dataset} using the same inversion code but different weights of the Stokes components in the fitting algorithm produced a slightly better $\sgn$ map (with a slight better agreement with \ME{} of 88.7\% for the case in \sect{sdm_720_reverse}) but presented some minor inversion artifacts in localized regions.
On the other hand,  other RTE inversions (with different parameters as well as with a different inversion code) consistently gave worse results. 
In other words, applications of the \SDM{}, as of any other disambiguation method, are sensitive to the quality of RTE inversions \kd{of spectropolarimetric observations of both employed telescopes}, and are expected to improve as refined inversion strategies will be developed. 

Second, the \SDM{} is based on a pixel-by-pixel comparison of observations with different resolutions and from different vantage points.
This implies that the employed remapping procedure (see \sect{implementation}) must attain sub-pixel accuracy. 
Hence, while a co-registration of the images is not required by the \SDM{} \textit{per se}, it is unlikely that the accuracy of the pointing information of both instruments can consistently reach such a precision, and co-registration will always be necessary.
The accuracy of our remapping procedure was tested using different, independently-developed routines, which produced the same co-registration shifts within a fraction of a \HRT{}-pixel.
\kdtwo
{\Fig{cnt_720_cc} proves that residual co-registration errors are 
within $\pm2$ pixels.} 
Therefore, residual co-registration errors cannot be the origin of the double-stripe in $\dBn$ that is discussed in \sect{sdm_720_reverse} (see in particular \fig{reverse_720_Bn}b), which is of the order of 8 pixels in width.
The 8-pixels shift is also larger than the amplitude of interpolation errors that could be expected by the difference in spatial resolution between \HRT{} and \HMI{} (which is a factor 2.6).
This was verified by using different numerical schemes for the interpolation involved in the co-registration and re-mapping procedures.
On the grounds of such tests and considerations, we tend to exclude co-registration inaccuracies as the possible origin of disambiguation errors, and of the double-stripe structure in $\dBn$ in particular.

Third, the \hmiME{} series that is used as input to the \SDM{} in \sect{sdm_720_reverse} is the result of a relatively long average (of about 20 minutes, but of co-registered and properly de-rotated images) in comparison with the integration time of the \HRT{} dataset (60\,s), and one may wonder if that difference could ultimately generate the shift seen in $\dBn$.
However, \sect{sdm_45_reverse_los} shows that similar results are obtained using the two series \hmiME{} and \hmiBlos{}, with the latter providing slightly better agreement with the results from applying \ME{} (possibly because of the better co-temporality of the employed datasets).
In a similar spirit, we have applied the \SDM{} using the \texttt{hmi.ME\_90s}, which is a series similar to the \hmiME{} series but with a cadence of 90\,s instead of 720\,s. 
Unfortunately, the \texttt{hmi.ME\_90s} series is not readily available for all observing times, in particular not for the March 17, 2022. 
However, a different \HRT{} observation \citep[March 7th, 2022, described in][]{Calchetti2023a,Sinjan2023} has also a corresponding \texttt{hmi.ME\_90s} dataset, which can be used as input to the \SDM{}. 
The observation date corresponds to when \SO{} crossed the Earth-Sun line.
While such a time is not suitable for stereoscopy ($\gamma=3.2^\circ$), it was still possible to study the resulting $\dBn$, and we found a very similar double-stripe structure as for  the \hmiME{} series presented in this paper.
Hence, from such tests, we can equally exclude that the amount of temporal averaging employed in the \HMI{} input magnetogram plays any substantial role in the generation of the apparent misalignment of $\Bn$ components, and of the related \SDM{} local inaccuracies. 
Incidentally, the presence of the double-stripe also at co-alignment (and with both \HRT{} and \HMI{} basically oriented along the solar north-south direction) also excludes that the remapping procedure plays any substantial role in generating the double-stripe.

\begin{figure}
 \setlength{\imsize}{\columnwidth}
 \centering
 \includegraphics[width=\imsize]{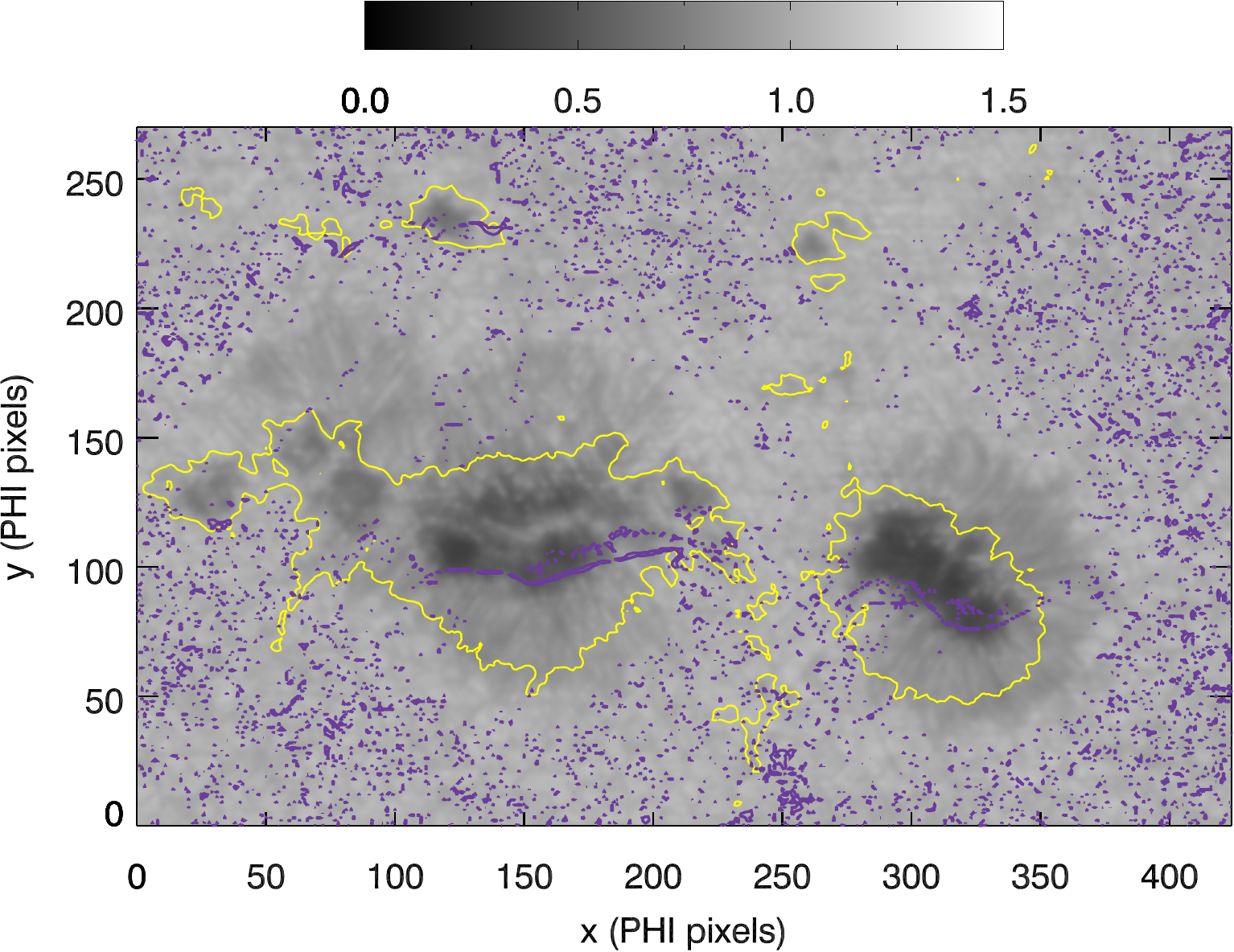}
 \\
 \caption{Continuum intensity of \HRT{} (same background image as \fig{cnt_720_vect}b) with $|\dBn|=0.85$ overlaid as \kd{purple} contour. The yellow solid curve represents the $|\Blos|$=400\,G isoline.}
\label{fig:light_bridge_720_vect}
\end{figure}
Fourth, a more subtle reason for a local mismatch between observations from different viewpoints might be related to the effective depth of formation of the line seen from the  two different vantage points.
In short, the signal registered by two instruments pointing at the same physical location on the Sun may not come from the same parcel of plasma, due to the different optical path that the light follows in the two different directions.
Such an effect is extensively discussed, and tested on numerical simulations, in Sect.~5 of \PaperI{}.
However, such an effect should decrease the more the observatories are aligned.
The above-mentioned observations at Sun-Earth line crossing offered the possibility to verify the importance of this effect, and we found no significant difference in the $\dBn$ double stripe structure. 
Similarly,  \fig{cnt_720_vect} shows the presence of a horizontal light-bridge across the AR polarities, qualitatively similar to the spatial distribution of the double-stripe in $\dBn$ of \fig{reverse_720_Bn}b. 
Due to the unmagnetized character of the plasma in the light bridge with respect to the surroundings, one could wonder if a similar optical-path effect is somehow affecting the \SDM{}  accuracy at the light-bridge location, thereby generating the double-stripe in $\dBn$.
However, \fig{light_bridge_720_vect} shows that there is a clear separation between the location of the light-bridge and the $\dBn$ double-stripe.
Hence, the apparent misalignment is not likely to be due to such an optical-path effect. 

Fifth, the  \HRT{} data that are used in this article were inverted using a standard version of the \HRT{} pipeline.
In particular, as clearly visible in \fig{cnt_720_vect}b, optical aberrations introduced by  thermal effects on the entrance window and geometrical distortions are not corrected for.
A procedure was recently developed to remove the former, which employs a point-spread function (PSF) obtained from phase-diversity (PD) analysis, convolved with the instrument's theoretical Airy disk.
We refer the reader to \cite{Kahil2022,Kahil2023} for more information on the PD analysis and the \HRT{} PSF, and to \citet{Sinjan2023,Calchetti2023a} for applications that implement such a procedure. 
Regarding the geometrical distortion, the procedure developed to correct for it is based on a spherical distortion model retrieved from on-ground data calibration only.
Concerning the \SDM{} application considered here, some moderate increase in the noise occurs as a result of the removal of the optical aberration, which requires to fine-tune the \ME{} disambiguation parameters for optimal performances, an effort that goes beyond the scope of this work.
As a result, the \SDM{} application is not improved by such corrections, while at the same time complicating the comparison with the \ME{} method, and is not included here.
However, geometrical distortions might indeed be one of the prime candidates for generating the mismatch in the $\Bn$-components associated to \SDM{} errors, and progress in this direction is expected to significantly improve \SDM{} results soon.

Sixth, \HMI{} observations are affected by an uncompensated Doppler line-shift due to the spacecraft radial velocity \citep{Hoeksema2014a, Couvidat2016, Schuck2016a}.
The employed \HMI{} datasets are taken when the \SDO{} radial velocity is about 2.4\,km\,s$^{-1}$, a velocity that has significant impact on the retrieved magnetic field parameters \citep[see in particular Fig.~18 in][]{Couvidat2016}. 
Such an effect may impact negatively the \SDM{} accuracy, in particular when the sign function $\sgn$ is computed using \eq{sdm_los} as in \sect{sdm_720_reverse_los} and \sect{sdm_45_reverse_los}, and even more so at relatively small $\gamma$ values such as that considered here.
\kd
{We suspect that such an effect may contribute to the apparently worse results that are obtained in the application of \eq{sdm_los} with respect to \eq{sdm_tr} to the particular \HRT{} dataset employed here.
}
At the time of writing, it was not possible to find \HMI{} observations that were co-temporal with \HRT{} data and recorded at a time when the effect of the radial \SDO{} velocity is minimal. 
However, new \SO{} observing campaigns are planned that will provide plenty of observations at different separation angles $\gamma$, and at times when \HMI{} radial velocity is close to zero.
Such forthcoming observations will allow to assess the ultimate effect of the \HMI{} radial velocity on \SDM{} accuracy. 

\subsection{The best---so far--observation-based disambiguation}\label{s:best}
\begin{figure*}
 \setlength{\imsize}{0.8\textwidth}
 \begin{center}
\labfigure{\imsize}{\figME/fig11a}{0.03\imsize} {0.70\imsize}{a)}
 \vspace{5pt} \\
\labfigure{\imsize}{\figME/fig11b}{0.03\imsize} {0.70\imsize}{b)}
 \end{center}
 \caption{
  Best---so far---observation-based \SDM{}-disambiguated vector magnetogram.
  The \HRT{} vector magnetogram is disambiguated using \HMI{} information from the (ambiguous) \hmiME{} data series and a combination of both \eq{sdm_tr} and \eq{sdm_los}, see \sect{best} for details.
   \kd{Panel a)} The background image represents $\Blos$ 
\kd{on the \HRT{} image plane,} saturated at $\pm1500$\,G in greyscale, and red/blue arrows represent the transverse field at positive/negative $\Blos$, respectively.
 The 400\,G-isoline of $|\Blos|$ is drawn as a \kd{yellow} solid line.
 \kd{Panel b)  same as panel a but in remapped Stonyhurst coordinates, with the magnetic field reprojected in radial, poloidal and toroidal components.
 In this panel, the background image  represents the radial component $B_r$ saturated at $\pm1500$\,G in greyscale, and red/blue arrows represent the horizontal field at positive/negative $B_r$, respectively.
 The 400\,G-isoline of $|B_r|$ is drawn as a yellow solid line.}
}
\label{fig:reverse_720_vect_mgm_combi}
\end{figure*}
Finally, improvements will come from more systematic  applications of the \SDM{} itself.
As already studied in \PaperI{}, the geometrical equivalence of \eqs{sdm_tr}{sdm_los} allows for combining the two: in each pixel, one is free to choose for the \SDM{} disambiguation the equation that has a higher level of reliability at that specific location.
Such a decision requires practical criteria based on measurable quantities. 

For instance, one can prescribe that in, each pixel, the equation that is applied is the one that yields a value of $\sgn$ that is closer to the nominal value, $\pm1$. 
By applying such a criterion to the data in \sect{sdm_720_reverse}, we indeed found an improvement of the overall \SDM{} disambiguation, with a $\sgn$ map that is, by construction, closer to the nominal values.
The resulting \SDM{}-disambiguated vector magnetogram is shown in \fig{reverse_720_vect_mgm_combi}a, which shows the best observation-based disambiguation obtained so far. 
We notice in particular that the inconsistencies around \kdtwo{\PIL{}} in the penumbral areas (orange rectangles in \fig{reverse_720_vect_tr}b) are for the most part removed, as well as a reduced effect of the errors related to the double-stripe in $\dBn$ (in the areas highlighted by the \kd{black} rectangles in \fig{reverse_720_vect_tr}b).   
\kd
{Similar conclusions can be drawn by comparing the disambiguated vector magnetogram in heliographic projection obtained in this case, \fig{reverse_720_vect_mgm_combi}b, with the correspondent magnetograms in \fig{reverse_720_vect_mgm}b (discussed in \sect{sdm_720_reverse_tr}) and \fig{app_helio}a,b (discussed in \sects{sdm_720_reverse_los}{sdm_45_reverse_los}, respectively). 
Again, the combined application of \eqs{sdm_tr}{sdm_los} yields a very significant reduction of areas of where the \SDM{} disambiguation is suspicious, and no unexpected reversal of the radial component is present.
} 
\Fig{reverse_720_ME0}\kd{c shows} that this combined disambiguation strategy also yields better agreement with \ME{} (86.6\%), especially in the eastern part of the strong field area (where \eq{sdm_los} is selected). 

In short, the development of a quantitative criterion to combine \eqs{sdm_tr}{sdm_los} is a promising way to further improve the accuracy and reliability of the \SDM{}. 
However, before attempting a general formulation of the required decision criteria, some of the systematic errors discussed above should first be removed.

\section{Conclusions}\label{s:conclusions}
We applied the \SDM{} to \HRT{} observations for the first time, demonstrating its viability for application to measured data beyond the proof-of-concept presented in \PaperI{}. 
This first test employs a dataset that was selected from the limited existing sample of \HRT{} observations in the first months of the science phase of \SO{}. 
Despite the fact that the available datasets are not yet completely favorable for optimum performance of the \SDM{}, we provide here the first ever observation-based, ambiguity-resolved magnetogram (\fig{reverse_720_vect_mgm_combi}). 
The intrinsic $180^\circ$ ambiguity of vector magnetograms has for the first time been removed solely thanks to stereoscopic observations, without making any assumption on the properties of the photospheric magnetic field. 
We have proven the feasibility of the \SDM{} approach in real applications, with overall very promising results. 
\kd{The best disambiguation, both in quantitative (\ie{} the sign function $\sgn$) as well qualitative (smoothness and expected orientation of the transverse field component) terms, is obtained by choosing in each pixel separately which one of the two geometrically equivalent \SDM{} formulae, \eq{sdm_tr} and \eq{sdm_los}, is closer to the nominal value, $\sgn=\pm1$.}

There is nonetheless room for improvement, as localized areas of errors are also found. 
Such areas are associated with measurable diagnostic quantities and we present in this paper several strategies for improvement.   

In  addition, we also present a preliminary comparison of \SDM{} results with a standard, single-viewpoint disambiguation method (the \ME{} method by \cite{Metcalf1994}).
Since standard methods are approximate insofar as they require to make assumptions about the properties of the  magnetic field, this is not a test for the \SDM{}.
However, the comparison is instructive (\fig{reverse_720_ME0}c):  differences between the two disambiguations are visible at some locations close to the \kdtwo{\PIL{} of $\Blos$} in penumbral areas, 
which are very interesting for further studies.
A very characteristic area of disagreement in the core of the sunspots was identified to be likely due to residual uncorrected geometrical distortions in \HRT{}. 
It is expected that the currently underway improvement in the reduction of \HRT{} data will greatly help to reduce such artifacts.

In conclusion, the \SDM{} fulfills the need for tools to exploit \SolO{} potentials for novel science, namely the stereoscopic disambiguation of vector magnetograms based on observed data only.   
In this way, unbiased benchmarking of single-view vector magnetograms, as well as the investigation of fundamental solar properties, such as a disambiguation-independent estimation of photospheric currents that are injected in upper coronal layers, are becoming possible.

\begin{acknowledgements}
\kd{We wish to thanks the referee for their competent and constructive comments that helped us improve the article.}
The authors thank Marco Stangalini for providing the co-registration routine, Bill Thompson for the development of the WCS routines that are part of the SolarSoft analysis package, and Pradeep Chitta for his insightful comments.
EP acknowledges financial support from the French national space agency (CNES) through the APR program. 
EP was also supported by the French Programme National PNST of CNRS/INSU co-funded by CNES and CEA.
Solar Orbiter is a space mission of international collaboration between ESA and NASA, operated by ESA. We are grateful to the ESA SOC and MOC teams for their support. The German contribution to SO/PHI is funded by the BMWi through DLR and by MPG central funds. 
The Spanish contribution is funded by AEI/MCIN/10.13039/501100011033/ and European Union ``NextGenerationEU/PRTR'' (RTI2018-096886-C5,  PID2021-125325OB-C5,  PCI2022-135009-2,PCI2022-135029-2) and ERDF ``A way of making Europe''; ''Center of Excellence Severo Ochoa'' awards to IAA-CSIC (SEV-2017-0709, CEX2021-001131-S); and a Ram\'on y Cajal fellowship awarded to DOS. The French contribution is funded by CNES.
The HMI data are courtesy of NASA/SDO and the HMI science team.
\end{acknowledgements}

 \bibliographystyle{aa} 


\begin{thebibliography}{45}
\expandafter\ifx\csname natexlab\endcsname\relax\def\natexlab#1{#1}\fi

\bibitem[{Albert {et~al.}(2023)Albert, Krivova, Hirzberger, Solanki, Vacas,
  Su\'arez, Jorge, Appourchaux, Alvarez-Herrero, Rodr\'iguez, Gandorfer,
  Gutierrez-Marques, Kahil, Kolleck, del Toro~Iniesta, Woch, Fiethe,
  P\'erez-Grande, Kilders, Jim\'enez, Rubio, Calchetti1, Carmona, Feller,
  Fernandez-Rico, Fern\'andez-Medina, Parejo, Blesa, Gizon, Grauf, Heerlein,
  Korpi-Lagg, Lange, Jim\'enez, Maue, Meller, M\"uller, Nakai, Schmidt, Schou,
  Sinjan, Staub, Strecker, Torralbo, \& Valori}]{Albert2023}
Albert, K., Krivova, N.~A., Hirzberger, J., {et~al.} 2023, submitted to \aap

\bibitem[{{Aulanier} {et~al.}(2013){Aulanier}, {D{\'e}moulin}, {Schrijver},
  {Janvier}, {Pariat}, \& {Schmieder}}]{Aulanier2013}
{Aulanier}, G., {D{\'e}moulin}, P., {Schrijver}, C.~J., {et~al.} 2013, \aap,
  549, A66

\bibitem[{Borrero {et~al.}(2011)Borrero, Tomczyk, Kubo, Socas-Navarro, Schou,
  Couvidat, \& Bogart}]{Borrero2011}
Borrero, J.~M., Tomczyk, S., Kubo, M., {et~al.} 2011, \solphys, 273, 267

\bibitem[{Calchetti {et~al.}(2023)Calchetti, Stangalini, Jafarzadeh, Valori,
  Albert, Albelo~Jorge, Alvarez-Herrero, Appourchaux, Balaguer~Jim{\'e}nez,
  Bellot~Rubio, Blanco~Rodr{\'\i}guez, Feller, Gandorfer, Germerott, Gizon,
  Guerrero, Gutierrez-Marques, Hirzberger, Kahil, Kolleck, Korpi-Lagg,
  Moreno~Vacas, Orozco~Su{\'a}rez, P{\'e}rez-Grande, Sanchis~Kilders, Schou,
  Sch{\"u}hle, Sinjan, Solanki, Staub, Strecker, del Toro~Iniesta, Volkmer, \&
  Woch}]{Calchetti2023a}
Calchetti, D., Stangalini, M., Jafarzadeh, S., {et~al.} 2023, \aap, 674, A109

\bibitem[{Couvidat {et~al.}(2012)Couvidat, Rajaguru, Wachter,
  Sankarasubramanian, Schou, \& Scherrer}]{Couvidat2012}
Couvidat, S., Rajaguru, S.~P., Wachter, R., {et~al.} 2012, \solphys, 278, 217

\bibitem[{Couvidat {et~al.}(2016)Couvidat, Schou, Hoeksema, Bogart, Bush,
  Duvall, Liu, Norton, \& Scherrer}]{Couvidat2016}
Couvidat, S., Schou, J., Hoeksema, J.~T., {et~al.} 2016, \solphys, 291, 1887

\bibitem[{{DeRosa} {et~al.}(2015){DeRosa}, {Wheatland}, {Leka}, {Barnes},
  {Amari}, {Canou}, {Gilchrist}, {Thalmann}, {Valori}, {Wiegelmann},
  {Schrijver}, {Malanushenko}, {Sun}, \& {R{\'e}gnier}}]{deRosa2015}
{DeRosa}, M.~L., {Wheatland}, M.~S., {Leka}, K.~D., {et~al.} 2015, \apj, 811,
  107

\bibitem[{{Forbes} {et~al.}(2006){Forbes}, {Linker}, {Chen}, {Cid}, {K{\'o}ta},
  {Lee}, {Mann}, {Miki{\'c}}, {Potgieter}, {Schmidt}, {Siscoe}, {Vainio},
  {Antiochos}, \& {Riley}}]{Forbes2006}
{Forbes}, T.~G., {Linker}, J.~A., {Chen}, J., {et~al.} 2006, Space Sci.~Rev.,
  123, 251

\bibitem[{Freeland \& Handy(2012)}]{Freeland2012}
Freeland, S.~L. \& Handy, B.~N. 2012, SolarSoft: Programming and data analysis
  environment for solar physics, Astrophysics Source Code Library, record
  ascl:1208.013

\bibitem[{Gandorfer {et~al.}(2018)Gandorfer, Grauf, Staub, Bischoff, Woch,
  Hirzberger, Solanki, {\'A}lvarez-Herrero, Parejo, Schmidt, Volkmer,
  Appourchaux, \& del Toro~Iniesta}]{Gandorfer2018}
Gandorfer, A.~M., Grauf, B., Staub, J., {et~al.} 2018, in Society of
  Photo-Optical Instrumentation Engineers (SPIE) Conference Series, Vol. 10698,
  Space Telescopes and Instrumentation 2022: Optical, Infrared, and Millimeter
  Wave, 1403--1415

\bibitem[{{Gary} \& {Hagyard}(1990)}]{Gary1990}
{Gary}, G.~A. \& {Hagyard}, M.~J. 1990, \solphys, 126, 21

\bibitem[{{Hoeksema} {et~al.}(2014){Hoeksema}, {Liu}, {Hayashi}, {Sun},
  {Schou}, {Couvidat}, {Norton}, {Bobra}, {Centeno}, {Leka}, {Barnes}, \&
  {Turmon}}]{Hoeksema2014a}
{Hoeksema}, J.~T., {Liu}, Y., {Hayashi}, K., {et~al.} 2014, \solphys, 289, 3483

\bibitem[{Kahil {et~al.}(2023)Kahil, Gandorfer, Hirzberger, Calchetti, Sinjan,
  Valori, Solanki, Van~Noort, Albert, Albelo~Jorge, Alvarez-Herrero,
  Appourchaux, Bellot~Rubio, Blanco~Rodr{\'\i}guez, Feller, Fiethe, Germerott,
  Gizon, Guerrero, Gutierrez-Marques, Kolleck, Korpi-Lagg, Michalik,
  Moreno~Vacas, Orozco~Su{\'a}rez, P{\'e}rez-Grande, Sanchis~Kilders, Schou,
  Sch{\"u}hle, Staub, Strecker, del Toro~Iniesta, Volkmer, \& Woch}]{Kahil2023}
Kahil, F., Gandorfer, A., Hirzberger, J., {et~al.} 2023, \aap, 675, A61

\bibitem[{Kahil {et~al.}(2022)Kahil, Gandorfer, Hirzberger, Orozco~Su{\'a}rez,
  Albert, Albelo~Jorge, Appourchaux, {\'A}lvarez-Herrero,
  Blanco~Rodr{\'\i}guez, Germerott, Guerrero, Gutierrez~Marquez, Sinjan,
  Calchetti, Kolleck, Solanki, del Toro~Iniesta, Volkmer, Woch, Fiethe,
  G{\'o}mez~Cama, P{\'e}rez-Grande, Sanchis~Kilders, Balaguer~Jim{\'e}nez,
  Bellot~Rubio, Carmona, Deutsch, Fernandez-Rico, Fern{\'a}ndez-Medina,
  Garc{\'\i}a~Parejo, Gasent~Blesa, Gizon, Grauf, Heerlein, Korpi-Lagg, Lange,
  L{\'o}pez~Jim{\'e}nez, Maue, Meller, Michalik, Moreno~Vacas, M{\"u}ller,
  Nakai, Schmidt, Schou, Sch{\"u}hle, Staub, Strecker, Torralbo, \&
  Valori}]{Kahil2022}
Kahil, F., Gandorfer, A., Hirzberger, J., {et~al.} 2022, in Society of
  Photo-Optical Instrumentation Engineers (SPIE) Conference Series, Vol. 12180,
  Space Telescopes and Instrumentation 2022: Optical, Infrared, and Millimeter
  Wave, ed. L.~E. {Coyle}, S.~{Matsuura}, \& M.~D. {Perrin}, 121803F

\bibitem[{LaBonte(2004)}]{LaBonte2004}
LaBonte, B. 2004, \solphys, 221, 191

\bibitem[{Leka \& Barnes(2012)}]{Leka2012}
Leka, K.~D. \& Barnes, G. 2012, \solphys, 277, 89

\bibitem[{{Leka} {et~al.}(2009){Leka}, {Barnes}, {Crouch}, {Metcalf}, {Gary},
  {Jing}, \& {Liu}}]{Leka2009}
{Leka}, K.~D., {Barnes}, G., {Crouch}, A.~D., {et~al.} 2009, \solphys, 260, 83

\bibitem[{Leka {et~al.}(2022)Leka, Wagner, Gri{\~n}{\'o}n-Mar{\'\i}n, Bommier,
  \& Higgins}]{Leka2022}
Leka, K.~D., Wagner, E.~L., Gri{\~n}{\'o}n-Mar{\'\i}n, A.~B., Bommier, V., \&
  Higgins, R. E.~L. 2022, \solphys, 297, 121

\bibitem[{Li \& Long(2023)}]{Li2023}
Li, D. \& Long, D.~M. 2023, \apj, 944, 8

\bibitem[{{Lites}(2000)}]{Lites2000}
{Lites}, B.~W. 2000, Rev.~Geophys., 38, 1

\bibitem[{Liu {et~al.}(2022)Liu, Gri{\~n}{\'o}n-Mar{\'\i}n, Hoeksema, Norton,
  \& Sun}]{Liu2022}
Liu, Y., Gri{\~n}{\'o}n-Mar{\'\i}n, A.~B., Hoeksema, J.~T., Norton, A.~A., \&
  Sun, X. 2022, \solphys, 297, 17

\bibitem[{{Mart{\'\i}nez Pillet}(2007)}]{Martinez-Pillet2007}
{Mart{\'\i}nez Pillet}, V. 2007, in ESA Special Publication, Vol. 641, Second
  Solar Orbiter Workshop, ed. E.~{Marsch}, K.~{Tsinganos}, R.~{Marsden}, \&
  L.~{Conroy}, 27

\bibitem[{Metcalf(1994)}]{Metcalf1994}
Metcalf, T.~R. 1994, \solphys, 155, 235

\bibitem[{{Metcalf} {et~al.}(2006){Metcalf}, {Leka}, {Barnes}, {Lites},
  {Georgoulis}, {Pevtsov}, {Balasubramaniam}, {Gary}, {Jing}, {Li}, {Liu},
  {Wang}, {Abramenko}, {Yurchyshyn}, \& {Moon}}]{Metcalf2006}
{Metcalf}, T.~R., {Leka}, K.~D., {Barnes}, G., {et~al.} 2006, \solphys, 237,
  267

\bibitem[{M{\"u}ller {et~al.}(2020)M{\"u}ller, St.~Cyr, Zouganelis, Gilbert,
  Marsden, Nieves-Chinchilla, Antonucci, Auch{\`e}re, Berghmans, Horbury,
  Howard, Krucker, Maksimovic, Owen, Rochus, Rodriguez-Pacheco, Romoli,
  Solanki, Bruno, Carlsson, Fludra, Harra, Hassler, Livi, Louarn, Peter,
  Sch{\"u}hle, Teriaca, del Toro~Iniesta, Wimmer-Schweingruber, Marsch, Velli,
  De~Groof, Walsh, \& Williams}]{Mueller2020}
M{\"u}ller, D., St.~Cyr, O.~C., Zouganelis, I., {et~al.} 2020, \aap, 642, A1

\bibitem[{Orozco~Su{\'a}rez \& Del Toro~Iniesta(2007)}]{OrozcoSuarez2007}
Orozco~Su{\'a}rez, D. \& Del Toro~Iniesta, J.~C. 2007, \aap, 462, 1137

\bibitem[{Pesnell {et~al.}(2012)Pesnell, Thompson, \& Chamberlin}]{Pesnell2012}
Pesnell, W.~D., Thompson, B.~J., \& Chamberlin, P.~C. 2012, \solphys, 275, 3

\bibitem[{Pietarila {et~al.}(2013)Pietarila, Bertello, Harvey, \&
  Pevtsov}]{Pietarila2013}
Pietarila, A., Bertello, L., Harvey, J.~W., \& Pevtsov, A.~A. 2013, \solphys,
  282, 91

\bibitem[{Rochus {et~al.}(2020)Rochus, Auch{\`e}re, Berghmans, Harra, Schmutz,
  Sch{\"u}hle, Addison, Appourchaux, Aznar~Cuadrado, Baker, Barbay, Bates,
  BenMoussa, Bergmann, Beurthe, Borgo, Bonte, Bouzit, Bradley, B{\"u}chel,
  Buchlin, B{\"u}chner, Cab{\'e}, Cadiergues, Chaigneau, Chares, Choque~Cortez,
  Coker, Condamin, Coumar, Curdt, Cutler, Davies, Davison, Defise, Del~Zanna,
  Delmotte, Delouille, Dolla, Dumesnil, D{\"u}rig, Enge, Fran{\c{c}}ois,
  Fourmond, Gillis, Giordanengo, Gissot, Green, Guerreiro, Guilbaud, Gyo,
  Haberreiter, Hafiz, Hailey, Halain, Hansotte, Hecquet, Heerlein, Hellin,
  Hemsley, Hermans, Hervier, Hochedez, Houbrechts, Ihsan, Jacques,
  J{\'e}r{\^o}me, Jones, Kahle, Kennedy, Klaproth, Kolleck, Koller, Kotsialos,
  Kraaikamp, Langer, Lawrenson, Le~Clech', Lenaerts, Liebecq, Linder, Long,
  Mampaey, Markiewicz-Innes, Marquet, Marsch, Matthews, Mazy, Mazzoli, Meining,
  Meltchakov, Mercier, Meyer, Monecke, Monfort, Morinaud, Moron, Mountney,
  M{\"u}ller, Nicula, Parenti, Peter, Pfiffner, Philippon, Phillips, Plesseria,
  Pylyser, Rabecki, Ravet-Krill, Rebellato, Renotte, Rodriguez, Roose, Rosin,
  Rossi, Roth, Rouesnel, Roulliay, Rousseau, Ruane, Scanlan, Schlatter, Seaton,
  Silliman, Smit, Smith, Solanki, Spescha, Spencer, Stegen, Stockman, Szwec,
  Tamiatto, Tandy, Teriaca, Theobald, Tychon, van Driel-Gesztelyi, Verbeeck,
  Vial, Werner, West, Westwood, Wiegelmann, Willis, Winter, Zerr, Zhang, \&
  Zhukov}]{Rochus2020}
Rochus, P., Auch{\`e}re, F., Berghmans, D., {et~al.} 2020, \aap, 642, A8

\bibitem[{Rouillard {et~al.}(2020)Rouillard, Pinto, Vourlidas, De~Groof,
  Thompson, Bemporad, Dolei, Indurain, Buchlin, Sasso, Spadaro, Dalmasse,
  Hirzberger, Zouganelis, Strugarek, Brun, Alexand~re, Berghmans, Raouafi,
  Wiegelmann, Pagano, Arge, Nieves-Chinchilla, Lavarra, Poirier, Amari, Aran,
  Andretta, Antonucci, Anastasiadis, Auch{\`e}re, Bellot~Rubio, Nicula, Bonnin,
  Bouchemit, Budnik, Caminade, Cecconi, Carlyle, Cernuda, Davila, Etesi,
  Espinosa~Lara, Fedorov, Fineschi, Fludra, G{\'e}not, Georgoulis, Gilbert,
  Giunta, Gomez-Herrero, Guest, Haberreiter, Hassler, Henney, Howard, Horbury,
  Janvier, Jones, Kozarev, Kraaikamp, Kouloumvakos, Krucker, Lagg, Linker,
  Lavraud, Louarn, Maksimovic, Maloney, Mann, Masson, M{\"u}ller, {\"O}nel,
  Osuna, Orozco~Suarez, Owen, Papaioannou, P{\'e}rez-Su{\'a}rez,
  Rodriguez-Pacheco, Parenti, Pariat, Peter, Plunkett, Pomoell, Raines,
  Riethm{\"u}ller, Rich, Rodriguez, Romoli, Sanchez, Solanki, St~Cyr, Straus,
  Susino, Teriaca, del Toro~Iniesta, Ventura, Verbeeck, Vilmer, Warmuth, Walsh,
  Watson, Williams, Wu, \& Zhukov}]{Rouillard2020}
Rouillard, A.~P., Pinto, R.~F., Vourlidas, A., {et~al.} 2020, \aap, 642, A2

\bibitem[{Scherrer {et~al.}(2012)Scherrer, Schou, Bush, Kosovichev, Bogart,
  Hoeksema, Liu, Duvall, Zhao, Title, Schrijver, Tarbell, \&
  Tomczyk}]{Scherrer2012a}
Scherrer, P.~H., Schou, J., Bush, R.~I., {et~al.} 2012, \solphys, 275, 207

\bibitem[{Schou {et~al.}(2012)Schou, Scherrer, Bush, Wachter, Couvidat,
  Rabello-Soares, Bogart, Hoeksema, Liu, Duvall, Akin, Allard, Miles, Rairden,
  Shine, Tarbell, Title, Wolfson, Elmore, Norton, \& Tomczyk}]{Schou2012}
Schou, J., Scherrer, P.~H., Bush, R.~I., {et~al.} 2012, \solphys, 275, 229

\bibitem[{Schuck {et~al.}(2016)Schuck, Antiochos, Leka, \&
  Barnes}]{Schuck2016a}
Schuck, P.~W., Antiochos, S.~K., Leka, K.~D., \& Barnes, G. 2016, \apj, 823,
  101

\bibitem[{{Semel} \& {Skumanich}(1998)}]{Semel1998}
{Semel}, M. \& {Skumanich}, A. 1998, \aap, 331, 383

\bibitem[{{Sinjan} {et~al.}(2023){Sinjan}, {Calchetti}, {Hirzberger}, {Kahil},
  {Valori}, {Solanki}, {Albert}, {Albelo Jorge}, {Alvarez-Herrero},
  {Appourchaux}, {Bellot Rubio}, {Blanco Rodr{\'\i}guez}, {Feller},
  {Gandorfer}, {Germerott}, {Gizon}, {G{\'o}mez Cama}, {Guerrero},
  {Gutierrez-Marques}, {Kolleck}, {Korpi-Lagg}, {Michalik}, {Moreno Vacas},
  {Orozco Su{\'a}rez}, {P{\'e}rez-Grande}, {Sanchis Kilders}, {Balaguer
  Jim{\'e}nez}, {Schou}, {Sch{\"u}hle}, {Staub}, {Strecker}, {del Toro
  Iniesta}, {Volkmer}, \& {Woch}}]{Sinjan2023}
{Sinjan}, J., {Calchetti}, D., {Hirzberger}, J., {et~al.} 2023, \aap, 673, A31

\bibitem[{Sinjan {et~al.}(2022)Sinjan, Calchetti, Hirzberger,
  Orozco~Su{\'a}rez, Albert, Albelo~Jorge, Appourchaux, Alvarez-Herrero,
  Blanco~Rodr{\'\i}guez, Gandorfer, Germerott, Guerrero, Gutierrez~Marquez,
  Kahil, Kolleck, Solanki, del Toro~Iniesta, Volkmer, Woch, Fiethe,
  G{\'o}mez~Cama, P{\'e}rez-Grande, Sanchis~Kilders, Balaguer~Jim{\'e}nez,
  Bellot~Rubio, Carmona, Deutsch, Fernandez-Rico, Fern{\'a}ndez-Medina,
  Garc{\'\i}a~Parejo, Gasent~Blesa, Gizon, Grauf, Heerlein, Korpi-Lagg, Lange,
  L{\'o}pez~Jim{\'e}nez, Maue, Meller, Michalik, Moreno~Vacas, M{\"u}ller,
  Nakai, Schmidt, Schou, Sch{\"u}hle, Staub, Strecker, Torralbo, \&
  Valori}]{Sinjan2022}
Sinjan, J., Calchetti, D., Hirzberger, J., {et~al.} 2022, in Society of
  Photo-Optical Instrumentation Engineers (SPIE) Conference Series, Vol. 12189,
  Society of Photo-Optical Instrumentation Engineers (SPIE) Conference Series,
  121891J

\bibitem[{Solanki {et~al.}(2020)Solanki, del Toro~Iniesta, Woch, Gandorfer,
  Hirzberger, Alvarez-Herrero, Appourchaux, Mart{\'\i}nez~Pillet,
  P{\'e}rez-Grande, Sanchis~Kilders, Schmidt, G{\'o}mez~Cama, Michalik,
  Deutsch, Fernandez-Rico, Grauf, Gizon, Heerlein, Kolleck, Lagg, Meller,
  M{\"u}ller, Sch{\"u}hle, Staub, Albert, Alvarez~Copano, Beckmann, Bischoff,
  Busse, Enge, Frahm, Germerott, Guerrero, L{\"o}ptien, Meierdierks,
  Oberdorfer, Papagiannaki, Ramanath, Schou, Werner, Yang, Zerr, Bergmann,
  Bochmann, Heinrichs, Meyer, Monecke, M{\"u}ller, Sperling,
  {\'A}lvarez~Garc{\'\i}a, Aparicio, Balaguer~Jim{\'e}nez, Bellot~Rubio,
  Cobos~Carracosa, Girela, Hern{\'a}ndez~Exp{\'o}sito, Herranz, Labrousse,
  L{\'o}pez~Jim{\'e}nez, Orozco~Su{\'a}rez, Ramos, Barandiar{\'a}n, Bastide,
  Campuzano, Cebollero, D{\'a}vila, Fern{\'a}ndez-Medina, Garc{\'\i}a~Parejo,
  Garranzo-Garc{\'\i}a, Laguna, Mart{\'\i}n, Navarro, N{\'u}{\~n}ez~Peral,
  Royo, S{\'a}nchez, Silva-L{\'o}pez, Vera, Villanueva, Fourmond, de~Galarreta,
  Bouzit, Hervier, Le~Clec'h, Szwec, Chaigneau, Buttice, Dominguez-Tagle,
  Philippon, Boumier, Le~Cocguen, Baranjuk, Bell, Berkefeld, Baumgartner,
  Heidecke, Maue, Nakai, Scheiffelen, Sigwarth, Soltau, Volkmer,
  Blanco~Rodr{\'\i}guez, Domingo, Ferreres~Sabater, Gasent~Blesa,
  Rodr{\'\i}guez~Mart{\'\i}nez, Osorno~Caudel, Bosch, Casas, Carmona, Herms,
  Roma, Alonso, G{\'o}mez-Sanjuan, Piqueras, Torralbo, Fiethe, Guan, Lange,
  Michel, Bonet, Fahmy, M{\"u}ller, \& Zouganelis}]{Solanki2020}
Solanki, S.~K., del Toro~Iniesta, J.~C., Woch, J., {et~al.} 2020, \aap, 642,
  A11

\bibitem[{Solanki {et~al.}(2015)Solanki, del Toro~Iniesta, Woch, Gandorfer,
  Hirzberger, Schmidt, Appourchaux, \& Alvarez-Herrero}]{Solanki2015}
Solanki, S.~K., del Toro~Iniesta, J.~C., Woch, J., {et~al.} 2015, in IAU
  Symposium, Vol. 305, Polarimetry, ed. K.~N. {Nagendra}, S.~{Bagnulo},
  R.~{Centeno}, \& M.~{Jes{\'u}s Mart{\'\i}nez Gonz{\'a}lez}, 108--113

\bibitem[{Solanki \& Montavon(1993)}]{Solanki1993}
Solanki, S.~K. \& Montavon, C. A.~P. 1993, \aap, 275, 283

\bibitem[{Stenflo(1985)}]{Stenflo1985}
Stenflo, J.~O. 1985, \solphys, 100, 189

\bibitem[{Sun(2013)}]{Sun2013}
Sun, X. 2013, arXiv e-prints, arXiv:1309.2392

\bibitem[{{Thompson}(2006)}]{Thompson2006}
{Thompson}, W.~T. 2006, \aap, 449, 791

\bibitem[{Valori {et~al.}(2022)Valori, L{\"o}schl, Stansby, Pariat, Hirzberger,
  \& Chen}]{Valori2022}
Valori, G., L{\"o}schl, P., Stansby, D., {et~al.} 2022, \solphys, 297, 12

\bibitem[{Volkmer {et~al.}(2012)Volkmer, Bosch, Feger, Gomez, Heidecke,
  Schmidt, Scheiffelen, Sigwarth, \& Soltau}]{Volkmer2012}
Volkmer, R., Bosch, J., Feger, B., {et~al.} 2012, in Society of Photo-Optical
  Instrumentation Engineers (SPIE) Conference Series, Vol. 8442, Space
  Telescopes and Instrumentation 2012: Optical, Infrared, and Millimeter Wave,
  ed. M.~C. {Clampin}, G.~G. {Fazio}, H.~A. {MacEwen}, \& J.~{Oschmann},
  Jacobus~M., 84424P

\bibitem[{Zouganelis {et~al.}(2020)Zouganelis, De~Groof, Walsh, Williams,
  M{\"u}ller, St~Cyr, Auch{\`e}re, Berghmans, Fludra, Horbury, Howard, Krucker,
  Maksimovic, Owen, Rodr{\'\i}guez-Pacheco, Romoli, Solanki, Watson, Sanchez,
  Lefort, Osuna, Gilbert, Nieves-Chinchilla, Abbo, Alexandrova, Anastasiadis,
  Andretta, Antonucci, Appourchaux, Aran, Arge, Aulanier, Baker, Bale,
  Battaglia, Bellot~Rubio, Bemporad, Berthomier, Bocchialini, Bonnin, Brun,
  Bruno, Buchlin, B{\"u}chner, Bucik, Carcaboso, Carr, Carrasco-Bl{\'a}zquez,
  Cecconi, Cernuda~Cangas, Chen, Chitta, Chust, Dalmasse, D'Amicis, Da~Deppo,
  De~Marco, Dolei, Dolla, Dudok~de Wit, van Driel-Gesztelyi, Eastwood,
  Espinosa~Lara, Etesi, Fedorov, F{\'e}lix-Redondo, Fineschi, Fleck, Fontaine,
  Fox, Gandorfer, G{\'e}not, Georgoulis, Gissot, Giunta, Gizon,
  G{\'o}mez-Herrero, Gontikakis, Graham, Green, Grundy, Haberreiter, Harra,
  Hassler, Hirzberger, Ho, Hurford, Innes, Issautier, James, Janitzek, Janvier,
  Jeffrey, Jenkins, Khotyaintsev, Klein, Kontar, Kontogiannis, Krafft,
  Krasnoselskikh, Kretzschmar, Labrosse, Lagg, Landini, Lavraud, Leon, Lepri,
  Lewis, Liewer, Linker, Livi, Long, Louarn, Malandraki, Maloney,
  Martinez-Pillet, Martinovic, Masson, Matthews, Matteini, Meyer-Vernet,
  Moraitis, Morton, Musset, Nicolaou, Nindos, O'Brien, Orozco~Suarez, Owens,
  Pancrazzi, Papaioannou, Parenti, Pariat, Patsourakos, Perrone, Peter, Pinto,
  Plainaki, Plettemeier, Plunkett, Raines, Raouafi, Reid, Retino, Rezeau,
  Rochus, Rodriguez, Rodriguez-Garcia, Roth, Rouillard, Sahraoui, Sasso, Schou,
  Sch{\"u}hle, Sorriso-Valvo, Soucek, Spadaro, Stangalini, Stansby, Steller,
  Strugarek, {\v{S}}tver{\'a}k, Susino, Telloni, Terasa, Teriaca,
  Toledo-Redondo, del Toro~Iniesta, Tsiropoula, Tsounis, Tziotziou, Valentini,
  Vaivads, Vecchio, Velli, Verbeeck, Verdini, Verscharen, Vilmer, Vourlidas,
  Wicks, Wimmer-Schweingruber, Wiegelmann, Young, \& Zhukov}]{Zouganelis2020}
Zouganelis, I., De~Groof, A., Walsh, A.~P., {et~al.} 2020, \aap, 642, A3

\end{thebibliography}

\begin{appendix}
\section{Vector magnetograms in heliografic projection}
\begin{figure}[H]
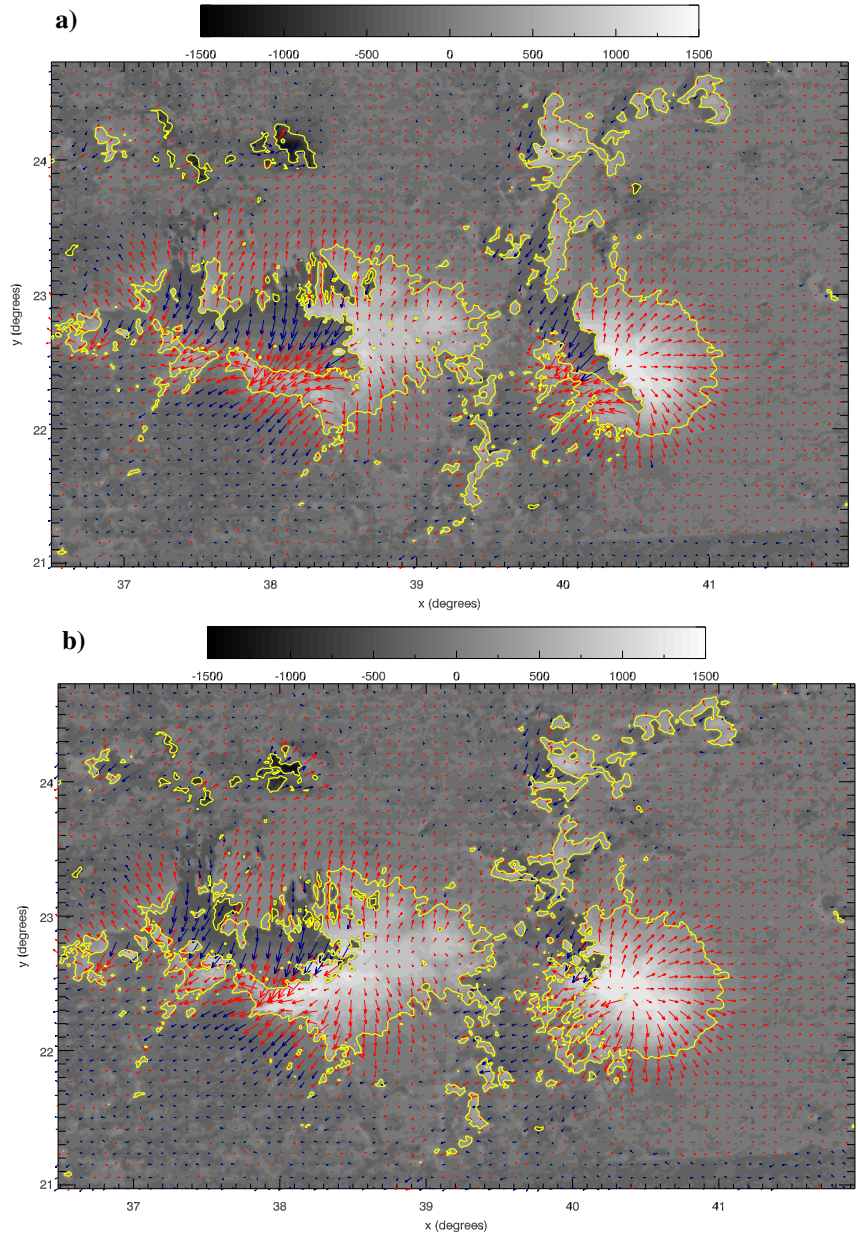

 \setlength{\imsize}{0.8\columnwidth}
 \begin{center}
 \labfigure {\imsize} {\figME/fig12a} {0.03\imsize} {0.70\imsize}{a)}
 \vspace{5pt} \\
 \labfigure {\imsize} {\figBlos/fig12b}{0.03\imsize}{0.70\imsize}{b)}
 \end{center}
 \caption{
\kd{
SDM-disambiguated vector magnetograms remapped in Stonyhurst coordinates, with the magnetic field reprojected in radial, poloidal and toroidal components, corresponding to the cases in \fig{reverse_720_vect_los}b  (panel a) and \fig{reverse_blos_vect}b (panel b), respectively.
 In both panels, the background image represents the radial component $B_r$ saturated at $\pm1500$\,G in greyscale, and red/blue arrows represent the horizontal field at positive/negative $B_r$, respectively.
 The 400\,G-isoline of $|B_r|$ is drawn as a yellow solid line.
}
 }
 \label{fig:app_helio}
\end{figure}
\end{appendix}
%
\end{document}